\documentclass[reprint,superscriptaddress,nofootinbib,floatfix,showkeys]{revtex4-1}

\usepackage{amsmath}
\usepackage{amssymb}
\usepackage{amsfonts}
\usepackage{amsthm}

\usepackage{graphicx}
\usepackage{color}
\usepackage{chemarr}
\usepackage{enumerate}
\usepackage[all]{xy}

\usepackage{bm}
\usepackage{bbm}

\usepackage{float} 
\usepackage{url}

\usepackage[labelfont=bf,labelsep=period,justification=raggedright]{caption}

\usepackage{xr}
\usepackage[version=3]{mhchem}

\newcommand{\ReactionR}[4]{#1 \ce{<=>[#3][#4]} #2}

\newcommand{\lp}{\left(}
\newcommand{\rp}{\right)}

\long\def\symbolfootnote[#1]#2{\begingroup \def\thefootnote{\fnsymbol{footnote}}\footnote[#1]{#2}\endgroup}

\newcommand{\mbf}[1]{\mathbf{#1}}
\newcommand{\trace}{\mathrm{trace}}
\newcommand{\abs}{\mathrm{abs}}
\newcommand{\mmol}{\mathrm{mmol/gDW/h}}
\newcommand{\VIt}{$ \textit{VI}(t)\,$}        
\newcommand{\VItt}{$\textit{VI}(t,t')\,$}
\newcommand{\VI}{\textit{VI}}

\newcommand{\nD}{\mbf{D}}
\newcommand{\nDp}{\bm{\mathcal{D}}}
\newcommand{\nDc}{\bm{\mathcal{D}}_{\mathrm{c}}}
\newcommand{\nDs}{\bm{\mathcal{D}}_{\mathrm{s}}}

\newcommand{\nM}[1]{\mbf{M}(\mathbf{v}^*_\mathrm{#1})}
\newcommand{\M}[1]{\mbf{M}_\mathrm{#1}}

\newcommand{\Adir}{\mbf{A}_{\mathrm{dir}}}
\newcommand{\nA}{\mbf{A}}
\newcommand{\nv}{\mbf{v}^*}

\newcommand{\C}[2]{C#1$({\mathrm{#2}})$}

\newcommand{\ie}{i.e.,~}

\begin{document}

\title{Flux-dependent graphs for metabolic networks}

\author{Mariano Beguerisse-D\'iaz} \email[]{beguerisse@maths.ox.ac.uk}
\affiliation{Department of
  Mathematics, Imperial College London, London, SW7 2AZ, UK}
\affiliation{Mathematical Institute, University
  of Oxford, Oxford OX2 6GG, UK}

\author{Gabriel Bosque} 
\affiliation{Institut Universitari d'Autom\`atica i Inform\`atica Industrial,
  Universitat Polit\`ecnica de Val\`encia, Cam\'i de Vera s/n, 46022
  Valencia, Spain.}

\author{Diego Oyarz\'un} \affiliation{Department of
  Mathematics, Imperial College London, London, SW7 2AZ, UK}

\author{Jes\'us Pic\'o} 
\affiliation{Institut Universitari d'Autom\`atica i Inform\`atica Industrial,
  Universitat Polit\`ecnica de Val\`encia, Cam\'i de Vera s/n, 46022
  Valencia, Spain.}

\author{Mauricio Barahona} \email[]{m.barahona@imperial.ac.uk}
\affiliation{Department of Mathematics, Imperial College London,
  London, SW7 2AZ, UK}

\begin{abstract}
Cells adapt their metabolic fluxes in response to changes in the environment. We present a framework for the systematic construction of flux-based graphs derived from organism-wide metabolic networks.  Our graphs encode the directionality of metabolic fluxes via edges that represent the flow of metabolites from source to target reactions.  The methodology can be applied in the absence of a specific biological context by modelling fluxes probabilistically, or can be tailored to different environmental conditions by incorporating flux distributions computed through constraint-based approaches such as Flux Balance Analysis. We illustrate our approach on the central carbon metabolism of {\it Escherichia coli} and on a metabolic model of human hepatocytes. The flux-dependent graphs under various environmental conditions and genetic perturbations exhibit systemic changes in their topological and community structure, which capture the re-routing of metabolic fluxes and the varying importance of specific reactions and pathways. By integrating constraint-based models and tools from network science, our framework allows the study of context-specific metabolic responses at a system level beyond standard pathway descriptions.
\end{abstract}

\maketitle

\onecolumngrid

\section{Introduction}

Metabolic reactions enable cellular function by converting nutrients
into energy, and by assembling macromolecules that sustain the cellular
machinery~\cite{Berg2002}. Cellular metabolism is usually thought of
as a collection of pathways comprising enzymatic reactions associated
with broad functional categories.  Yet metabolic reactions are highly
interconnected: enzymes convert multiple reactants into products with
other metabolites acting as co-factors; enzymes can catalyse several
reactions, and some reactions are catalysed by multiple enzymes,
and so on.  This enmeshed web of reactions is thus naturally amenable
to network analysis, an approach that has been successfully applied to
different aspects of cellular and molecular biology, e.g.,
protein-protein interactions~\cite{Thomas2003}, transcriptional
regulation~\cite{Alon2007}, or protein structure~\cite{Amor2014,amor2016prediction}.

Tools from graph theory~\cite{Newman2010} have previously been
applied to the analysis of structural properties of metabolic
networks, including their degree
distribution~\cite{Jeong2000,Wagner2001,Gleiss2001,Arita2004}, the
presence of metabolic roles~\cite{Guimera2005}, and their community
structure~\cite{Ravasz2002,Takemoto2013,Zhou2012,Cooper2010}.  A
central challenge, however, is that there are multiple ways to
construct a network from a metabolic
model~\cite{Palsson2006}.  For example, one can create a graph with
metabolites as nodes and edges representing the reactions that
transform one metabolite into
another~\cite{Ouzounis2000,Jeong2000,Wagner2001,Ma2003b}; a graph with
reactions as nodes and edges corresponding to the metabolites shared
among them~\cite{Ma2004,Vitkup2006,Samal2006}; or even a bipartite
graph with both reactions and metabolites as
nodes~\cite{Smart2008}. Importantly, the conclusions of
graph-theoretical analyses are highly dependent on the chosen graph
construction~\cite{Winterbach2013}.

A key feature of metabolic reactions is the directionality of flux:
metabolic networks contain both irreversible and reversible reactions,
and reversible reactions can change their direction depending on the
cellular and environmental contexts~\cite{Berg2002}.  Many of
the existing graph constructions, however, lead to undirected graphs that
disregard such directional information, which is central to metabolic
function~\cite{Palsson2006, Wagner2001}.  Furthermore, current graph
constructions are usually derived from the whole set of metabolic
reactions in an organism, and thus correspond to a generic
metabolic `blueprint' of the cell. However, cells switch specific
pathways `on' and `off' to sustain their energetic budget in different
environments~\cite{Sauer1999}. Hence, such blueprint graphs might not
capture the specific metabolic connectivity in a given environment,
thus limiting their ability to provide biological insights in
different growth conditions.

In this paper, we present a flux-based approach to construct metabolic
graphs that encapsulate the directional flow of metabolites produced
or consumed through enzymatic reactions.  The proposed graphs can be
tailored to incorporate flux distributions under different
environmental conditions. To introduce our approach, we proceed in two
steps.  We first define the \textit{Probabilistic Flux Graph} (PFG), a
weighted, directed graph with reactions as nodes, edges that represent
supplier-consumer relationships between reactions, and weights given
by the probability that a metabolite chosen at random is
produced/consumed by the source/target reaction.  This graph can be
used to carry out graph-theoretical analyses of organism-wide 
metabolic organisation independent of cellular context or environmental conditions.  
We then show that this formalism can be adapted seamlessly to construct the
{\it Metabolic Flux Graph} (MFG), a directed, environment-dependent,
graph with weights computed from Flux Balance Analysis
(FBA)~\cite{Orth2010a}, the most widespread method to study
genome-scale metabolic networks.

Our formulation addresses several drawbacks of current constructions
of metabolic graphs.  Firstly, in our flux graphs, an edge indicates
that metabolites are produced by the source reaction and consumed by
the target reaction, thus accounting for metabolic directionality and
the natural flow of chemical mass from reactants to products.
Secondly, the Probabilistic Flux Graph discounts naturally the
over-representation of pool metabolites (e.g., adenosine triphosphate
(ATP), nicotinamide adenine dinucleotide (NADH), protons, water, and
other co-factors) that appear in many reactions and tend to obfuscate
the graph connectivity. Our construction avoids the removal of pool
metabolites from the network, which can change the graph structure
drastically ~\cite{Ma2003,Croes2006,Silva2007,Kreimer2008,Samal2011}.
Finally, the Metabolic Flux Graph incorporates additional biological
information reflecting the effect of the environmental context into
the graph construction. In particular, since the weights in the MFG
correspond directly to fluxes (in units of mass per time), different
biological scenarios can be analysed using balanced fluxes (e.g., from
different FBA solutions) under different carbon sources and other
environmental perturbations~\cite{Palsson2006,Schuetz2007,Orth2010a, Rabinowitz2012}. 

After introducing the mathematical framework, we showcase our approach
with two examples. Firstly, in the absence of environmental context,
our analysis of the PFG of the core model of {\it Escherichia coli}
metabolism~\cite{Orth2010} reveals the importance of including
directionality and appropriate edge weights in the graph to understand
the modular organisation of metabolic sub-systems.  We then use FBA
solutions computed for several relevant growth conditions for {\it
  E. coli}, and show that the structure of the MFG changes
dramatically in each case (e.g., connectivity, ranking of reactions,
community structure), thus capturing the environment-dependent nature
of metabolism.  Secondly, we study a model of human hepatocyte
metabolism evaluated under different conditions for the wild-type and in a
mutation found in primary hyperoxaluria type 1, a rare metabolic
disorder~\cite{Pagliarini2016}, and show how the changes in network
structure of the MFGs reveal new information that is complementary to
the analysis of fluxes predicted by FBA.

\section{Results}

\subsection{Definitions and background}
Consider a metabolic network composed of $n$ metabolites $X_i$
($i=1,\ldots,n$) that participate in $m$ reactions
\begin{equation}
  R_j:\quad \ReactionR{\sum_{i=1}^n \alpha_{ij} X_i}{\sum_{i=1}^n
    \beta_{ij} X_i}{}{}\label{Eq:Reac},\quad j=1,2,\ldots,m,
\end{equation}
where $\alpha_{ij}$ and $\beta_{ij}$ are the stoichiometric
coefficients of species $i$ in reaction $j$. Let us denote the
concentration of metabolite $X_i$ at time $t$ as $x_i(t)$. 
We then define the $n$-dimensional vector of metabolite concentrations:
$\mbf{x}(t) = (x_1(t), \ldots, x_n(t))^T$ .  Each reaction takes place
with rate $v_j(\mbf{x}, t)$, measured in units of concentration per
time~\cite{Heinrich2012}. We compile these reaction rates in the 
$m$-dimensional vector: $\mbf{v}(t) = (v_1(t), \ldots, v_m(t))^T$.

\begin{figure}[tp]
\begin{center}
\includegraphics[width=1\textwidth]{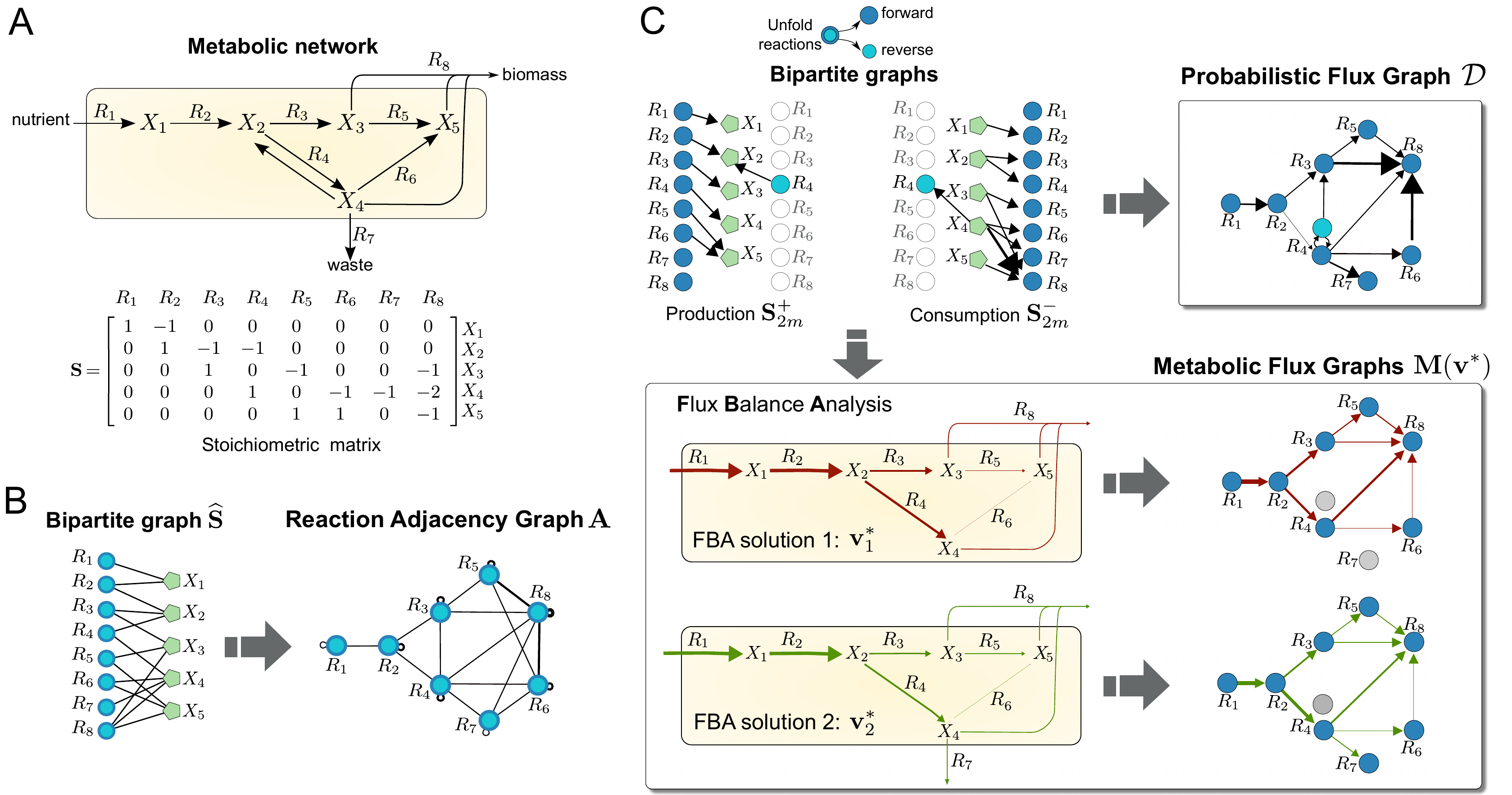} 
\end{center}
\caption{{\bf Graphs from metabolic networks.}  (A) Toy metabolic network describing nutrient uptake, biosynthesis of metabolic intermediates, secretion of waste products, and biomass production~\cite{Rabinowitz2012}.  The biomass reaction is $R_{8}:\,X_{3}+2X_{4}+X_{5}$.  (B) Bipartite graph associated with the boolean stoichiometric matrix $\widehat{\mbf{S}}$, and the Reaction Adjacency Graph (RAG)~\cite{Palsson2006} with adjacency matrix $\nA=\widehat{\mbf{S}}^{T}\widehat{\mbf{S}}$. The undirected edges of $\nA$ indicate the number of shared metabolites among reactions.  (C) The Probabilistic Flux Graph (PFG) $\nDp$ and two Metabolic Flux Graphs (MFG) $\nM{}$ constructed from the consumption and production stoichiometric matrices~\eqref{eq:s2rinout}. Note that the reversible reaction $R_4$ is unfolded into two nodes.  The PFG in Eq.~\eqref{eq:Dp} is a directed graph with weights representing the probability that the source reaction produces a metabolite consumed by the target reaction.  The MFGs in Eq.~\eqref{eq:Mv} are constructed from two different Flux Balance Analysis solutions ($\mathbf{v^*_1}$ and $\mathbf{v^*_2}$) obtained by optimising a biomass objective function under different flux constraints representing different environmental or cellular contexts (see Sec.~SI~2 in the Supplementary Information for details).  The weighted edges of the MFGs represent mass flow from source to target reactions in units of metabolic flux.  The computed FBA solutions translate into different connectivity in the resulting MFGs.}
\label{fig:ToyModel_networks}
\end{figure}

The mass balance of the system can then be represented compactly by
the system of ordinary differential equations
\begin{align}
\label{eq:metmod}
   \dot{\mbf{x}} &= \mbf{Sv},
\end{align}
where the $n \times m$ matrix $\mbf{S}$ is the stoichiometric matrix with
entries $S_{ij}=\beta_{ij}-\alpha_{ij}$, i.e., the net number of
$X_{i}$ molecules produced (positive $S_{ij}$) or consumed (negative
$S_{ij}$) by the $j$-th reaction.
Figure~\ref{fig:ToyModel_networks}A shows a toy example of a metabolic
network including nutrient uptake, biosynthesis of metabolic
intermediates, secretion of waste products, and biomass production~\cite{Rabinowitz2012}.

There are several ways to construct a graph for a given metabolic network
with stoichiometric matrix $\mbf{S}$.  A common approach~\cite{Palsson2006} is
to define the \textit{unipartite graph} with reactions as nodes and
$m \times m$ adjacency matrix
\begin{equation}
  {\nA} = \widehat{\mbf{S}}^{T} \, \widehat{\mbf{S}},
  \label{eq:A}
\end{equation}
where $\widehat{\mbf{S}}$ is the boolean version of $\mbf{S}$ (i.e.,
$\hat{S}_{ij} =1$ when $S_{ij}\neq0$ and $\hat{S}_{ij} =0$ otherwise).
This is the \textit{Reaction Adjacency Graph} (RAG), in which two reactions (nodes)
are connected if they share metabolites, either as reactants or
products. Self-loops represent the total number of metabolites
that participate in a reaction (Fig.~\ref{fig:ToyModel_networks}B).

Though widely studied~\cite{Wagner2001, Palsson2006}, the RAG has
known limitations and overlooks key aspects of the
connectivity of metabolic networks. The RAG does not
distinguish between forward and backward fluxes, nor does it
incorporate the irreversibility of reactions (by construction ${\nA}$
is a symmetric matrix).  Furthermore, the structure of ${\nA}$ is
dominated by the large number of edges introduced by pool
metabolites that appear in
many reactions, such as water, ions or enzymatic cofactors.  Computational schemes have been introduced to
mitigate the bias caused by pool metabolites~\cite{Croes2006}, but these do not follow from biophysical considerations and need manual calibration. Finally, the construction of the graph ${\nA}$ from 
is not easily extended to incorporate the effect of 
environmental changes.

\subsection{Metabolic graphs that incorporate flux directionality and biological context}

To address the limitations of the reaction adjacency graph
${\nA}$, we propose a graph formulation that
follows from a flux-based perspective. 
To construct our graph, we unfold each reaction into two separate directions (forward and reverse) 
and redefine the links between reaction nodes to reflect producer-consumer relationships.
Specifically, two reactions are connected if one produces a metabolite that is consumed
by the other.  As shown below, this definition leads to graphs that
naturally account for the reversibility of reactions, and allows for the
seamless integration of biological contexts modelled through FBA.

Inspired by matrix formulations of chemical reaction network
kinetics~\cite{Chellaboina2009}, we 
rewrite the reaction rate vector $\mbf{v}$ as: 
\begin{align*}
\mbf{v} & :=
\mbf{v^{+}} - \mbf{v^{-}} = \mbf{v^{+}} -
\textrm{diag}\left(\mbf{r}\right)\mbf{v^{-}},
\end{align*}
where $\mbf{v^{+}}$ and $\mbf{v^{-}}$ are non-negative vectors
containing the forward and backward reaction rates, respectively. Here 
the $m\times m$ matrix $\textrm{diag}\left(\mbf{r}\right)$ contains 
$\mbf{r}$ in its main diagonal, and
$\mbf{r}$ is the $m$-dimensional reversibility vector with components
$r_j = 1$ if reaction $R_j$ is reversible and $r_j = 0$ if it is
irreversible.  
With these definitions, we can rewrite the metabolic model in
Eq.~\eqref{eq:metmod} as:
\begin{align}
  \mbf{\dot{x}}=\mbf{S v} =
 \underbrace{
  \begin{bmatrix} \mbf{S} & -\mbf{S}\end{bmatrix} 
  \begin{bmatrix}\mbf{I}_m & 0 \\ 
    0 & \mathrm{diag}\left(\mbf{r}\right)
  \end{bmatrix}
 }_{\mbf{S}_{2m}}
  \begin{bmatrix}\mbf{v^{+}}\\
      \mbf{v^{-}}\end{bmatrix}
      : = \mbf{S}_{2m} \, \mbf{v}_{2m} \, ,  
  \label{eq:decomp} 
\end{align}
where $\mbf{v}_{2m} : = [ \mbf{v^{+}} \,\,\, \mbf{v^{-}} ]^T$
is the unfolded $2m$-dimensional vector of reaction rates,
$\mbf{I}_m$ is the $m\times m$ identity matrix, and we have defined
$\mbf{S}_{2m}$, the unfolded version of the stoichiometric matrix
of the $2m$ forward and reverse reactions.

\subsubsection{Probabilistic Flux Graph: a directional blueprint of metabolism}

The unfolding into forward and backward fluxes leads us to the definition of 
{\it production} and {\it consumption} stoichiometric matrices:
\begin{align}
  \begin{split}
   \text{Production:} \quad \mbf{S}_{2m}^{+} &= \frac{1}{2}
    \left( \abs\left(\mbf{S}_{2m}\right) + \mbf{S}_{2m}\right)\\
    \text{Consumption:} \quad \mbf{S}_{2m}^{-} &= \frac{1}{2}
    \left( \abs\left(\mbf{S}_{2m}\right) - \mbf{S}_{2m}
    \right),
  \end{split} \label{eq:s2rinout}
\end{align}
where $\abs\left(\mbf{S}_{2m}\right)$ is the matrix of 
absolute values of the corresponding entries of $\mbf{S}_{2m}$.
Note that each entry of the matrix $\mbf{S}_{2m}^{+}$, denoted
$s_{ij}^{+}$, gives the number of molecules of metabolite $X_{i}$
produced by reaction $R_{j}$. Conversely, the entries of
$\mbf{S}_{2m}^{-}$, denoted $s_{ij}^{-}$, correspond to the number of
molecules of metabolite $X_{i}$ consumed by reaction $R_{j}$.

Within our directional flux framework, it is natural to consider a purely
probabilistic description of producer-consumer relationships between
reactions, as follows.  Suppose we are given a stoichiometric matrix
$\mbf{S}$ without any additional biological information, such
as metabolite concentrations, reaction fluxes, or kinetic rates. In the
absence of such information, the probability that 
metabolite $X_k$ is produced by reaction $R_i$ and consumed by 
reaction $R_j$ is:
\begin{equation}
P\left(\text{a molecule of $X_k$ is produced by $R_i$ and consumed
    by $R_j$}\right) = \frac{s^+_{ki}}{w_k^+} \,
\frac{s^-_{kj}}{w_k^-},
\label{eq:prob}
\end{equation}
where $w_{k}^{+}=\sum_{h=1}^{2m} s_{kh}^{+}$ and
$w_{k}^{-}=\sum_{h=1}^{2m} s_{kh}^{-}$ are the total number of
molecules of $X_k$ produced and consumed by all reactions. Unlike
models in that rely on stochastic chemical
kinetics~\cite{gillespie1977}, the probabilities in
Eq.~\eqref{eq:prob} do not contain information on kinetic rate
constants, which are typically not available for genome-scale
metabolic models~\cite{Srinivasan2015}. In our formulation, the
relevant probabilities contain only the stoichiometric information
included in the matrix $\mbf{S}_{2m}$ and should not be confused with
the reaction propensity functions in Gillespie-type stochastic
simulations of biochemical systems.


We thus define the weight of the edge between reaction nodes $R_i$
 and $R_j$ as the probability that {\it any} metabolite chosen at
 random is produced by $R_i$ and consumed by $R_j$. Summing over all
 metabolites and normalizing, we obtain the edge weights of the adjacency
 matrix of the PFG: 
\begin{equation}
  \mathcal{D}_{ij} = \frac{1}{n}\sum_{k=1}^n\frac{s^+_{ki}}{w_k^+} \,
  \frac{s^-_{kj}}{w_k^-},
\label{eq:dp_ij}
\end{equation}
in which $\sum_{i,j} \mathcal{D}_{ij} = 1$ (i.e., the probability that
any metabolite is consumed/produced by any reaction is 1).  Rewritten
compactly in matrix form, we obtain the
\begin{align} 
  \textbf{Probabilistic Flux Graph (PFG):} \quad \quad \nDp &= \frac{1}{n} 
  \lp \mbf{W}_{+}^{\dagger} \mbf{S}_{2m}^{+} \rp^T 
  \lp \mbf{W}_{-}^{\dagger} \mbf{S}_{2m}^{-} \rp,
  \label{eq:Dp}
\end{align}
where $\mbf{W}_{+} ^\dagger= \text{diag}(\mbf{S}_{2m}^{+}
\mathbbm{1}_{2m})^\dagger$, $\mbf{W}_{-} ^\dagger=
\text{diag}(\mbf{S}_{2m}^{-} \mathbbm{1}_{2m})^\dagger$,
$\mathbbm{1}_{2m}$ is a vector of ones, and $\dagger$ denotes the
Moore-Penrose pseudoinverse.  
In Figure~\ref{fig:ToyModel_networks}C
we illustrate the creation of the PFG for a toy network. 
The PFG is a weighted, directed graph which encodes a blueprint of the
whole metabolic model, and provides a natural scaling of the contribution of pool
metabolites to flux transfer.  We remark that the PFG is distinct from directed analogues of 
the RAG constructed from boolean production and consumption stoichiometric matrices, 
as shown in Sec.~SI~1.

We now extend the construction of the PFG to accommodate
specific environmental contexts or growth conditions.

\subsubsection{Metabolic Flux Graphs: incorporating information of the biological context}

Cells adjust their metabolic fluxes to respond to the availability of
nutrients and environmental requirements. Flux Balance Analysis (FBA)
is a widely used method to predict environment-specific flux
distributions. FBA computes a vector of metabolic fluxes $\mbf{v}^{*}$
that maximise a cellular objective (e.g., biomass, growth or ATP
production). The FBA solution is obtained assuming steady state
conditions ($\mbf{\dot{x}}=0$ in Eq.~\eqref{eq:metmod}) 
subject to constraints that describe the availability
of nutrients and other extracellular compounds~\cite{Palsson2006}. The
core elements of FBA are briefly summarised in Section~\ref{sec:fba}.

To incorporate the biological information afforded by FBA solutions
into the structure of a metabolic graph, we again define the graph
edges in terms of production and consumptions fluxes.  Similarly to
Eq.~\eqref{eq:decomp}, we unfold the FBA solution vector $\nv$ into
forward and backward components: positive entries in the FBA
solution correspond to forward fluxes, negative entries in the
FBA solution correspond to backward fluxes. From the unfolded fluxes
\begin{equation*}
  \mbf{v}_{2m}^* = \begin{bmatrix}
    {\mbf{v}^{*}}^{+} \\
    {\mbf{v}^{*}}^{-}
  \end{bmatrix} =\frac{1}{2}\begin{bmatrix}
    \mathrm{abs}\lp\mbf{v}^{*}\rp + \mbf{v}^{*}\\
    \mathrm{abs}\lp\mbf{v}^{*}\rp - \mbf{v}^{*}
  \end{bmatrix},
\end{equation*}
we compute the vector of production and consumption fluxes as
\begin{align}
\mathbf{j}(\mathbf{v}^*) = \mbf{S}_{2m}^+ \mbf{v}_{2m}^* =
\mbf{S}_{2m}^- \mbf{v}_{2m}^*.	
\end{align}
The $k$-th entry of $\mathbf{j}(\mathbf{v}^*)$ is the flux at which
metabolite $X_k$ is produced and consumed, and the equality of the production and
consumption fluxes follows from the steady state condition, $\mbf{\dot{x}}=0$. 

To construct the flux graph, we define the weight of the edge
between reactions $R_i$ and $R_j$ as the \emph{total flux of
  metabolites produced by $R_i$ that are consumed by $R_j$}. Assuming
that the amount of metabolite produced by one reaction is distributed
among the reactions that consume it in proportion to their flux (and respecting
the stoichiometry), the flux of metabolite $X_k$ from reaction $R_i$ to $R_j$ is given by
\begin{align}
\text{Flux of $X_k$ from $R_i$ to $R_j$} = (\text{flux of $X_k$
  produced by $R_i$}) \times \lp \frac{\text{flux of $X_k$ consumed by
    $R_j$}}{\text{total consumption flux of $X_k$}} \rp.
          \label{eq:M-description}
\end{align} 
For example, if the total flux of metabolite $X_{k}$ is $10~\mmol$,
with reaction $R_{i}$ producing $X_k$ at a rate $1.5~\mmol$ and
reaction $R_j$ consuming $X_{k}$ at a rate $3.0~\mmol$, then the flux
of $X_k$ from $R_i$ to $R_j$ is $0.45~\mmol$. 

Summing~\eqref{eq:M-description} over all metabolites, we obtain the
edge weight relating reactions $R_i$ and $R_j$:
\begin{align}
  M_{ij}(\mathbf{v}^*) &= \sum_{k=1}^{n} s_{ki}^{+}{v}_{2m\, i}^* \times
  \left(\frac{s_{kj}^{-}{v}_{2m\, j}^*}{\sum_{j=1}^{2m}
    s_{kj}^{-}{v}_{2m\, j}^*}\right). \label{eq:mij}
\end{align}
In matrix form, these edge weights are collected into the adjacency matrix of the
 \begin{align}
  \textbf{Metabolic Flux Graph (MFG):} \quad \quad \nM{} &= 
\lp\mbf{S}_{2m}^{+}\mbf{V}^*\rp^T {\mbf{J}}_v^\dagger
      \lp\mbf{S}_{2m}^{-}\mbf{V}^*\rp,
      \label{eq:Mv}
\end{align}
where $\mbf{V}^{*} = \mathrm{diag}\lp\mbf{v}_{2m}^*\rp$,
$\mbf{J}_v = \mathrm{diag} \lp \mathbf{j}(\mathbf{v}^*)\rp$ and
$\dagger$ denotes the matrix pseudoinverse.  The MFG is a directed, weighted
graph with edge weights in units of $\mmol$.  Self-loops describe the metabolic flux of 
autocatalytic reactions, \ie those in which products are also reactants.

The MFG provides a versatile framework to create environment-specific
metabolic graphs from FBA solutions.  In Figure~\ref{fig:ToyModel_networks}C, 
we illustrate the creation of MFGs for a toy network under
different biological scenarios. In each case, an FBA solution is computed 
under a fixed uptake flux with the remaining fluxes constrained to account for
differences in the biological environment:  in scenario 1, the fluxes are
constrained to be strictly positive and no larger than the nutrient
uptake flux, while in scenario 2 we impose a positive lower bound on
reaction $R_7$.  Note how the MFG for scenario 2 displays an extra edge
between reactions $R_4$ and $R_7$, as well as distinct edge weights 
to scenario 1 (see Sec.~SI~2 for details). These differences illustrate 
how changes in the FBA solutions translate into
different graph connectivities and edge weights.

\subsection{Flux-based graphs of \textit{Escherichia coli} metabolism}

\begin{figure}[tp]
 \begin{center}
\includegraphics[width=\textwidth]{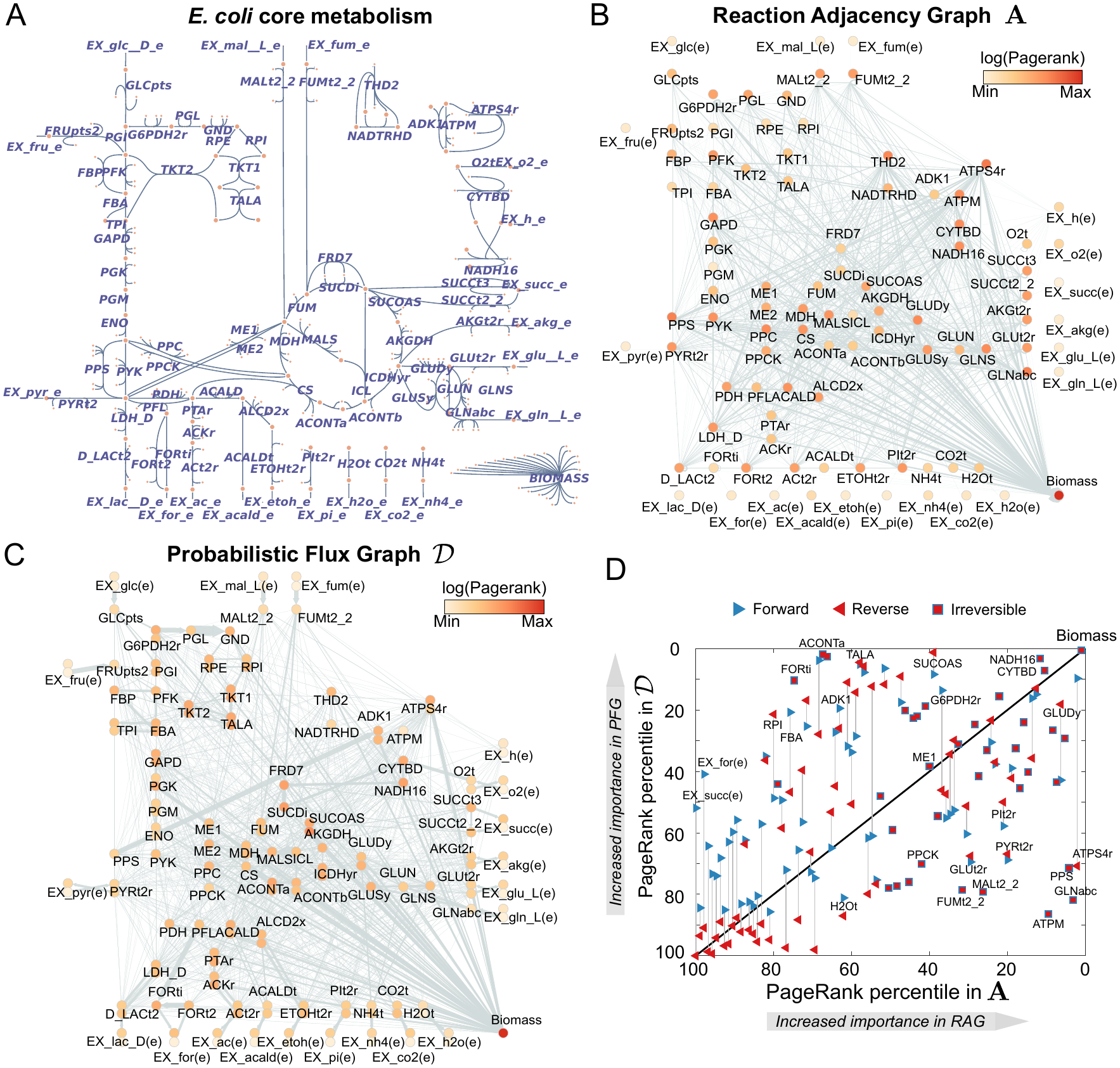}
 \end{center}
 \caption{\textbf{Graphs for the core metabolism of {\it Escherichia coli}.}  (A) Map of the {\it E. coli} core metabolic model created with the online tool Escher~\cite{Orth2010,King2015}.  (B) The standard Reaction Adjacency Graph ${\nA}$, as given by Eq.~\eqref{eq:A}. The nodes represent reactions; two reactions are linked by an undirected edge if they share reactants or products.  The nodes are coloured according to their PageRank score, a measure of their centrality (or importance) in the graph.  (C) The directed Probabilistic Flux Graph $\nDp$, as computed from Eq.~\eqref{eq:Dp}.  The reversible reactions are unfolded into two overlapping nodes (one for the forward reaction, one for the backward). The directed links indicate flow of metabolites produced by the source node and consumed by the target node.  The nodes are coloured according to their PageRank score.  (D) Comparison of PageRank percentiles of reactions in $\nA$ and $\nDp$.  Reversible reactions are represented by two triangles connected by a line; both share the same PageRank in $\nA$, but each has its own PageRank in $\nDp$.  Reactions that appear above (below) the diagonal have increased (decreased) PageRank in $\nDp$ as compared to $\nA$.}
 \label{fig:Model_A_Dnorm}
\end{figure}

To illustrate our framework, we construct and analyse the flux graphs of the well-studied core metabolic model of {\it
  E. coli}~\cite{Orth2010}. This model (Fig.~\ref{fig:Model_A_Dnorm}A)
contains 72 metabolites and 95 reactions, grouped into 11 pathways,
which describe the main biochemical routes in central carbon
metabolism~\cite{folch2015,Schuster2000,Schilling2000}. We provide a
Supplemental Spreadsheet with full details of the reactions and
metabolites in this model, as well as all the results presented below.

\subsubsection{The Probabilistic Flux Graph: the impact of directionality}

To examine the effect of flux directionality on the metabolic
graphs, we compare the Reaction Adjacency Graph (${\nA}$) and our proposed
Probabilistic Flux Graph ($\nDp$) for the same metabolic model in
Figure \ref{fig:Model_A_Dnorm}. The ${\nA}$ graph has 95 nodes and 
1,158 undirected edges, whereas the $\nDp$ graph has 154 nodes and 
1,604 directed and weighted edges. The
increase in node count is due to the unfolding of forward and backward
reactions into separate nodes. Unlike the $\nA$ graph, where edges
represent shared metabolites between two reactions, the directed edges
of the $\nDp$ graph represent the flow of metabolites from a source to
a target reaction. A salient feature of both graphs is their high
connectivity, which is not apparent from the traditional pathway
representation in Figure~\ref{fig:Model_A_Dnorm}A.

The effect of directionality 
becomes apparent 
when 
 comparing 
the importance of reaction nodes in both graphs (Figure~\ref{fig:Model_A_Dnorm}B--D)
, as measured with the 
PageRank score for node centrality~\cite{Page1999,Gleich2015}. 
The overall node hierarchy is maintained across both graphs: exchange reactions tend to have low PageRank centrality scores, core metabolic reactions
have high scores, and the biomass reaction has
the highest scores in both graphs. Yet we also observe substantial changes in specific reactions. For example, the reactions 
for ATP maintenance (ATPM, irreversible), phosphoenolpyruvate synthase (PPS, irreversible) and
ABC-mediated transport of L-glutamine (GLNabc, irreversible) drop from
being among the top 10\% most important reactions in the $\nA$ graph
to the bottom percentiles in the $\nDp$ graph. Conversely,
reactions such as aconitase A (ACONTa, irreversible), transaldolase
(TALA, reversible) and succinyl-CoA synthetase (SUCOAS, reversible),
and formate transport via diffusion (FORti, irreversible) gain
substantial importance in the $\nDp$ graph.  For instance, FORti is
the sole consumer of formate, which is produced by pyruvate formate
lyase (PFL), a reaction that is highly connected to the rest of the
network.  Importantly, in most of the reversible reactions, such as
ATP synthase (ATPS4r), there is a wide gap between the PageRank of the
forward and backward reactions, suggesting a marked asymmetry in the
importance of metabolic flows. 

\begin{figure}[tp]
 \begin{center}
   \includegraphics[width=\textwidth]{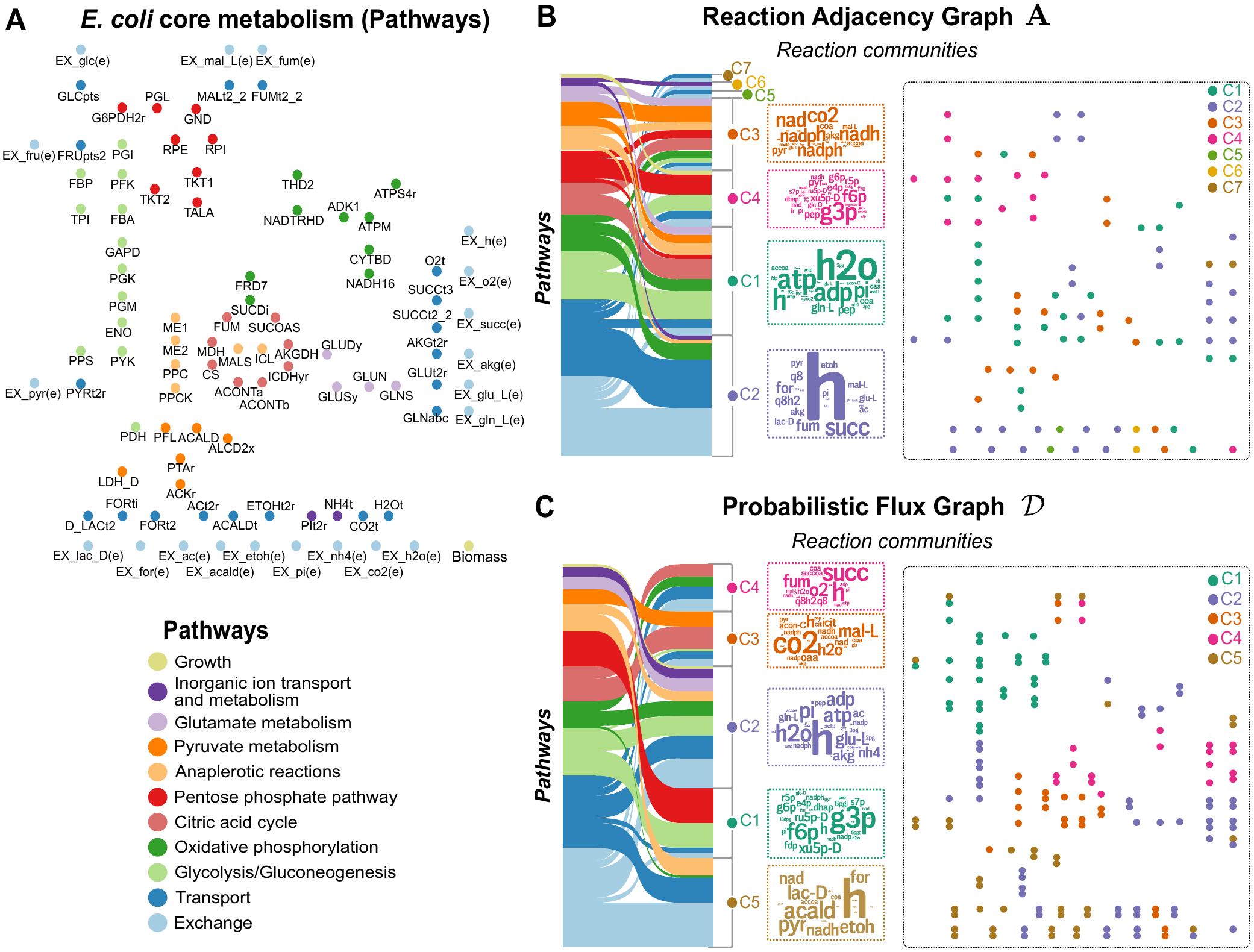}
 \end{center}
 \caption{{\bf Directionality and community structure of graphs for \textit{Escherichia coli} metabolism.}  (A) Reactions of the core model of {\it E. coli} metabolism grouped into eleven biochemical pathways, (B--C) Graphs ${\nA}$ and $\nDp$ from Fig.~\ref{fig:Model_A_Dnorm}B--C partitioned into communities computed with the Markov Stability method; for clarity, the graph edges are not shown.  The Sankey diagrams~\cite{Sankey1896,Rosvall2010} show the correspondence between biochemical pathways and the communities found in each graph. The word clouds contain the metabolites that participate in the reactions each community, and the word size is proportional to the number of reactions in which each metabolite participates.  }
 \label{fig:AvsD_comms}
\end{figure}

Community detection is frequently used in the
analysis of complex graphs: nodes are clustered into tightly related
communities that 
reveal the coarse-grained structure of the
graph, potentially at different levels of resolution~\cite{Girvan2002,
  Schaub2012, Lambiotte2014}. The community structure of
metabolic graphs 
has been the subject of multiple
analyses~\cite{Zhou2012,Ravasz2002,Girvan2002}.  However, most
community detection methods are 
applicable to undirected
graphs only, and thus fail to capture the directionality of the metabolic graphs we propose here. 
To account for graph directionality, we use
the Markov Stability community detection
framework~\cite{Delvenne2010,Delvenne2013, Lambiotte2014}, which
uses diffusion on graphs to detect groups of nodes where flows are
retained persistently across time scales. 
Markov Stability is ideally suited to find multi-resolution
community structure~\cite{Schaub2012} and can deal with both directed
and undirected graphs~\cite{Beguerisse2014, Lambiotte2014} (see
Sec.~\ref{sec:markov}). 
In the case of metabolic graphs, Markov Stability can 
reveal groups of reactions that are closely interlinked
by the flow of metabolites that they produce and consume.

Figure~\ref{fig:AvsD_comms} shows the difference between
the community structure of the undirected RAG and the directed PFG of
the core metabolism of \textit{E. coli}.  For the  $\nA$ graph, Markov Stability 
reveals a partition into seven communities
(Figure~\ref{fig:AvsD_comms}B, see also Sec.~SI~3), which
are largely dictated by the many edges created by shared pool metabolites. 
For example, community \C{1}{\nA} is mainly composed of reactions 
that consume or produce ATP and water. Yet, the biomass reaction (the largest
consumer of ATP) is not a member of \C{1}{\nA} because, in the standard 
$\nA$ graph construction, any connection involving ATP has equal weight.
Other communities in $\nA$ are also determined by pool metabolites,
e.g.  \C{2}{\nA} is dominated by H$^+$, and \C{3}{\nA} is dominated by
NAD$^+$ and NADP$^+$, as illustrated by word clouds of the
relative frequency of metabolites in the reactions within each community.  
The community structure in ${\nA}$ thus
reflects the limitations of the RAG construction due to the absence
of biological context and the large number of uninformative links
introduced by pool metabolites.

For the $\nDp$ graph, we found a robust partition into five communities
 (Figure~\ref{fig:AvsD_comms}C, see also Sec.~SI~3), which
comprise reactions related consistently through biochemical pathways.  
Community \C{1}{\nDp} contains the reactions in
the pentose phosphate pathway together with the first steps of
glycolysis involving D-fructose, D-glucose, or D-ribulose.  Community
\C{2}{\nDp} contains the main reactions that produce ATP from
substrate level as well as oxidative phosphorylation and the biomass
reaction.  Community \C{3}{\nDp} includes the core of the citric acid
cycle, anaplerotic reactions related to malate syntheses, as well as
the intake of cofactors such as CO$_{2}$.  Community \C{4}{\nDp}
contains reactions that are secondary sources of carbon (such as
malate and succinate), as well as oxidative phosphorilation reactions.
Finally, community \C{5}{\nDp} contains reactions that are part of the
pyruvate metabolism subsystem, as well as transport reactions for the
most common secondary carbon metabolites such as lactate, formate,
acetaldehyde and ethanol.  Altogether, the communities of the $\nDp$
graph reflect metabolite flows associated with specific cellular
functions, as a consequence of including flux directionality in the graph construction. 
As seen in Fig.~\ref{fig:AvsD_comms}C, the communities are no 
longer exclusively determined by pool metabolites (e.g., water is no longer dominant and
protons are spread among all communities).  For a more detailed
explanation and comparison of the communities found in the $\nA$ and
$\nDp$ graphs, see Section~SI~3. Full information 
about PageRank scores and communities is provided in the Supplementary Spreadsheet.

\subsubsection{ Metabolic Flux Graphs: the impact of growth conditions and
  biological context}

\begin{figure}[tp]
 \begin{center}
     \includegraphics[width=\textwidth]{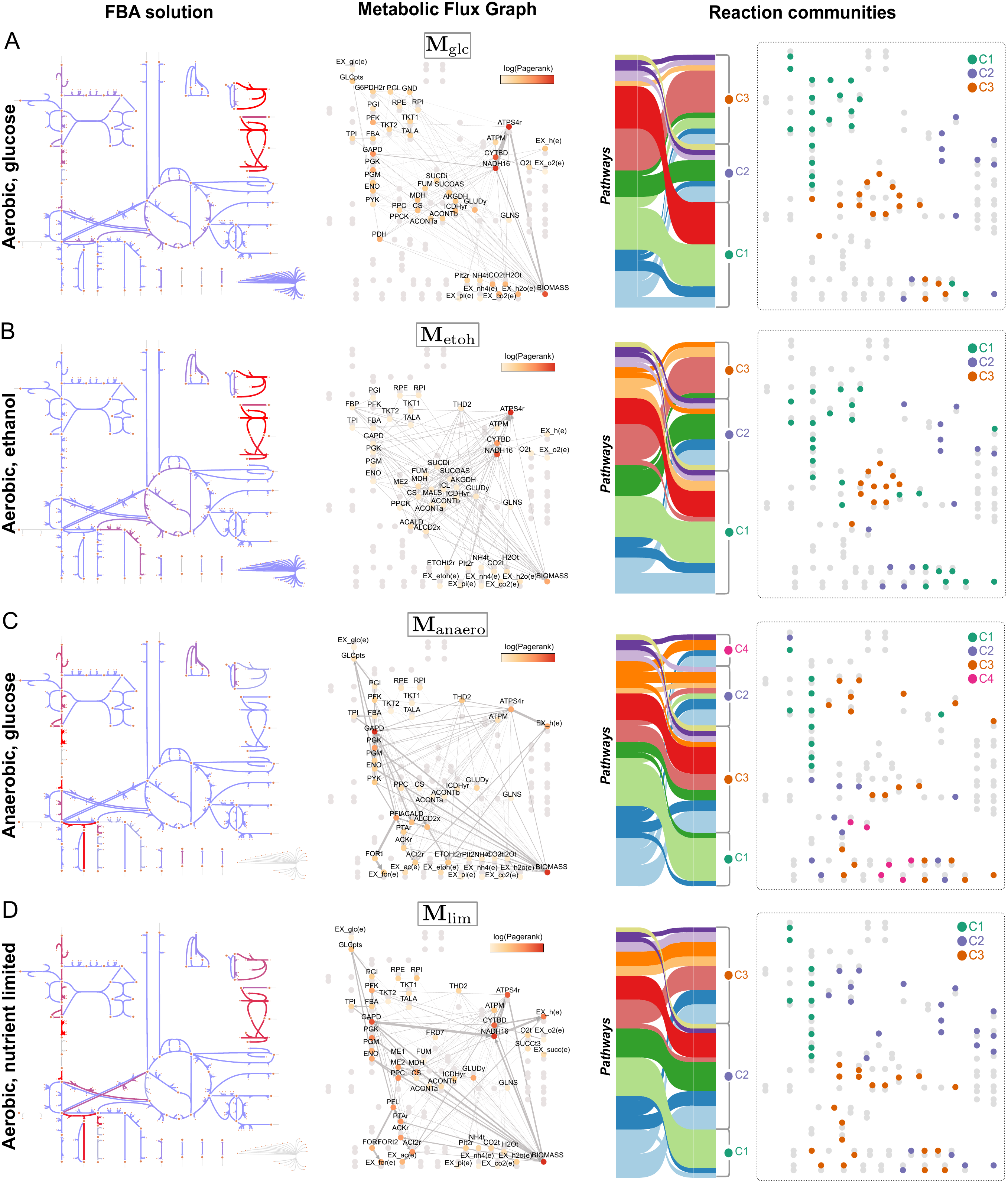} 
 \end{center}
 \caption{{\bf Metabolic Flux Graphs for \textit{Escherichia coli} in different growth conditions.} The MFGs are computed from Eq.~\eqref{eq:Mv} and the FBA solutions in four different environments: (A) aerobic growth in D-glucose, (B) aerobic growth in ethanol, (C) anaerobic growth in D-glucose, and (D) aerobic growth in D-glucose but with limited ammonium and phosphate.  Each subfigure shows: (left) flux map obtained with Escher~\cite{King2015}, where the increased red colour of the arrows indicates increased flux; (centre) Metabolic Flux Graph with nodes coloured according to their PageRank (zero flux reactions are in grey; thickness of connections proportional to fluxes); (right) community structure computed with the Markov Stability method together with Sankey diagrams showing the correspondence between biochemical pathways and MFG communities.}
 \label{fig:networksM}
\end{figure}

To incorporate the impact of environmental context
, we
construct Metabolic Flux Graphs in Eq.~\eqref{eq:Mv} using FBA solutions of
the core
model of {\it E. coli} metabolism in four relevant growth conditions:
aerobic growth in rich media with glucose;  aerobic growth in rich media with ethanol, 
anaerobic growth in glucose; and aerobic growth
in glucose but phosphate- and ammonium-limited.
The results in Figure~\ref{fig:networksM} 
show how changes in metabolite fluxes under different biological contexts have
a direct effect in the MFG.  Note that, in all cases, the MFGs have fewer nodes than the
blueprint graph $\nDp$ since the FBA solutions contain numerous
reactions with zero flux. 

Next we summarise how the changes in the community structure of the MFGs for the four
conditions reflect the distinct relationships of functional pathways 
in response to growth requirements.

\paragraph{Aerobic growth in D-glucose ($\M{glc}$).} 
We found a robust partition into three communities with
an intuitive biological interpretation (Fig.~\ref{fig:networksM}A and
Fig.~SI~2A).
\C{1}{\M{glc}} is the carbon-processing community,
comprising reactions that process carbon from D-glucose to pyruvate
including most of the glycolysis and pentose phosphate pathways,
together with related transport and exchange reactions. 
\C{2}{\M{glc}} harbours the bulk of reactions related to oxidative
phosphorylation and the production of energy in the cell, including
the electron transport chain of NADH dehydrogenase, cytochrome
oxidase, and ATP synthase, as well as transport reactions for
phosphate and oxygen intake and proton balance. \C{2}{\M{glc}}
also includes the growth reaction, consistent with ATP being
the main substrate for both the ATP maintenance (ATPM) requirement and
the biomass reaction in this growth condition.  Finally,
\C{3}{\M{glc}} contains reactions related to the citric acid cycle
(TCA) and the production of NADH and NADPH (\ie the cell's reductive
power), together with carbon intake routes
strongly linked to the TCA cycle, such as those starting from 
phosphoenolpyruvic acid (PEP).

\paragraph{Aerobic growth in ethanol ($\M{etoh}$).} 
The robust partition into three communities that we found for this 
scenario resembles the structure of $\M{glc}$ with subtle, yet important, differences
(Fig.~\ref{fig:networksM}B and
Fig.~SI~2B).  
Most salient are the differences in the carbon-processing community
\C{1}{\M{etoh}}, which reflects the switch
from D-glucose to ethanol as a carbon source.  
\C{1}{\M{etoh}} contains gluconeogenic
reactions (instead of glycolytic), due to the reversal of flux induced
by the change of carbon source, as well as anaplerotic reactions and
reactions related to glutamate metabolism.  In particular, the
reactions in this community are related to the production of precursors such as
PEP, pyruvate, 3-phospho-D-glycerate (3PG), glyceraldehyde-3-phosphate
(G3P), D-fructose-6-phosphate (F6P), and D-glucose-6-phosphate, all of
which are substrates for growth.  Consequently, the biomass reaction
is also grouped within \C{1}{\M{etoh}} due to the increased metabolic
flux of precursors relative to ATP production in this biological
scenario.  The other two reaction communities (energy-generation
\C{2}{\M{etoh}} and citric acid cycle \C{3}{\M{etoh}}) display less
prominent differences relative to the $\M{glc}$ graph, with additional
pyruvate metabolism and anaplerotic reactions as well as subtle
ascriptions of reactions involved in NADH/NADPH balance and the source
for acetyl-CoA.

\paragraph{Anaerobic growth in D-glucose ($\M{anaero}$).} 
The profound impact of the absence of oxygen on the metabolic balance
of the cell is reflected in drastic changes in the MFG (Fig.~\ref{fig:networksM}C and
Fig.~SI~2C). Both the connectivity and reaction communities in the MFG 
are starkly different from the aerobic scenarios, with a much diminished presence of oxidative
phosphorylation and the absence of the first two steps of the
electron transport chain (CYTBD and NADH16).  We found that
$\M{anaero}$ has a robust partition into four communities.
\C{1}{\M{anaero}} still contains carbon processing (glucose intake and
glycolysis), yet now decoupled from the pentose
phosphate pathway. \C{3}{\M{anaero}} includes the pentose phosphate pathway
grouped with the citric acid cycle (incomplete) and the biomass
reaction, as well as the growth precursors including alpha-D-ribose-5-phosphate (r5p),
D-erythrose-4-phosphate (e4p), 2-oxalacetate and NADPH.  The other two
communities are specific to the anaerobic context: \C{2}{\M{anaero}}
contains the conversion of PEP into formate (more than half of the
carbon secreted by the cell becomes formate~\cite{Sawers2005});
\C{4}{\M{anaero}} includes NADH production and consumption via
reactions linked to glyceraldehyde-3-phosphate dehydrogenase (GAPD).
 
\paragraph{Aerobic growth in D-glucose but limited phosphate and ammonium ($\M{lim}$).} 
Under growth-limiting conditions, we found a robust partition
into three communities (Fig.~\ref{fig:networksM}D and
Fig.~SI~2D).  The community
structure reflects \textit{overflow metabolism}~\cite{Vemuri2007},
which occurs when the cell takes in more carbon than it can process.
As a consequence, the excess carbon is secreted from the cell, leading
to a decrease in growth and a partial shutdown of the citric
acid cycle.  This is reflected in the reduced weight of the TCA
pathway and its grouping with the secretion routes
of acetate and formate within \C{3}{\M{lim}}. Hence, 
\C{3}{\M{lim}} comprises reactions that
are not strongly coupled in favourable growth conditions, yet
are linked together by metabolic responses to limited ammonium 
and phosphate.  Furthermore, the carbon-processing community \C{1}{\M{lim}} 
contains the glycolytic pathway, yet detached from the pentose phosphate pathway 
(as in $\M{anaero}$), highlighting its role in precursor formation. The
bioenergetic machinery, contained in community \C{2}{\M{lim}},
includes the pentose phosphate pathway, with a smaller role for the
electron transport chain (21.8\% of the total ATP as compared to
66.5\% in $\M{glc}$).


\begin{figure}[htp!]
 \begin{center}
      \includegraphics[width=\textwidth]{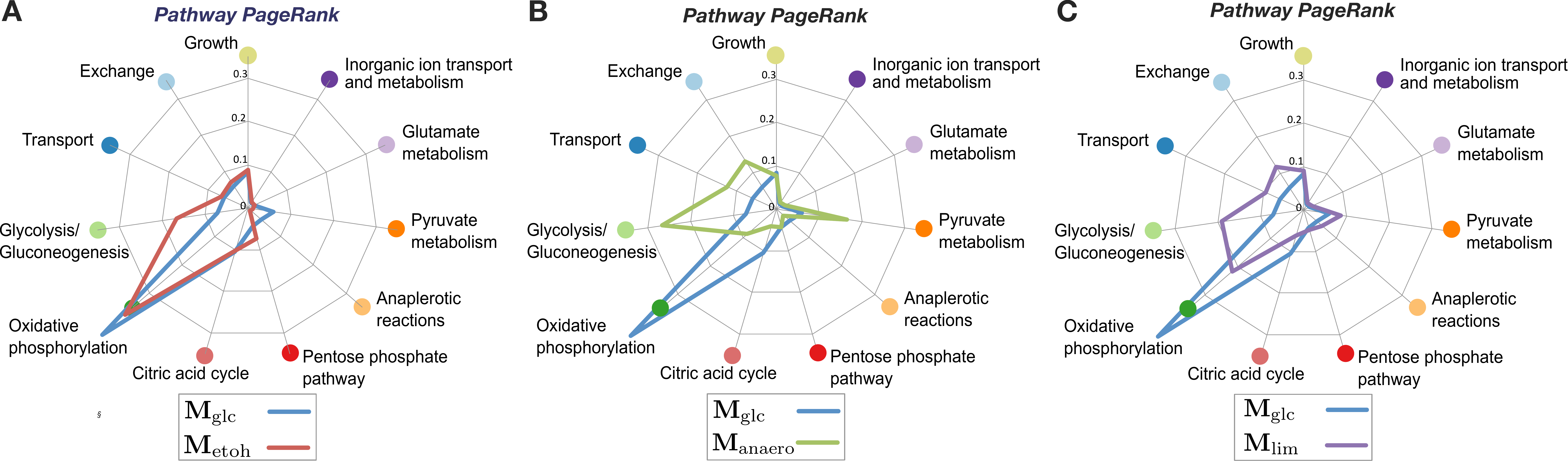} 
 \end{center}
 \caption{{\bf Pathway centrality (PageRank) computed from the MFG of different growth conditions.} The cumulative pathway PageRank reflects the relative importance of metabolic pathways in each MFG. Changes in pathway centrality indicate the overall rearrangement of fluxes within the pathways in response to environmental shifts: (A) from aerobic glucose-rich to aerobic ethanol-rich; (B) from aerobic glucose-rich to anerobic glucose-rich; (C) from aerobic glucose-rich to a similar medium with limited phosphate and ammonium. Variations in cumulative Pagerank highlight changes across most cellular pathways .}
 \label{fig:pagerank_pathway}
\end{figure}

In addition to the effect on community structure, 
Figure~\ref{fig:networksM} also shows the changes induced by the environment on the
MFG connectivity and relative importance of reactions, as measured by
their PageRank score. To provide a global snapshot of the effect of 
growth conditions on cellular metabolism, Figure~\ref{fig:pagerank_pathway} shows 
the cumulative PageRank of each pathway for each of the MFGs.  
The cumulative PageRank quantifies the relative importance of 
pathways, and how their importance changes upon environmental shifts. 
%
In aerobic growth, a shift from glucose to ethanol ($\M{glc} \to \M{etoh}$) as carbon source increases 
the importance of pyruvate metabolism and oxidative phosphorylation, while reducing the importance of the pentose phosphate pathway. A shift from aerobic to anaerobic growth in glucose ($\M{glc} \to \M{anaero}$) sees a large reduction in the importance of oxidative phosphorylation and the citric acid cycle, coupled with a large increase in the importance of gluconeogenesis, pyruvate metabolism, and transport and exchange reactions. The effect of growth-limiting conditions in aerobic growth under glucose ($\M{glc} \to \M{lim}$) is reflected on the increased importance of pyruvate metabolism and a reduction in the importance of oxidative phosphorylation, citric acid cycle, and the pentose phosphate pathway. The importance of transport and exchange reactions is also increased under limiting conditions. Such qualitative relations between growth conditions and the importance of specific pathways highlights the utility of the MFGs to characterise systemic metabolic changes in response to environmental conditions.

 
A more detailed discussion of the changes in pathways 
and reactions can be found in Section~SI~4 and
Fig.~SI~2 in the Supplementary Information, with full details of all
the results in the Supplemental Spreadsheet.

\subsubsection{Multiscale organisation of metabolic flux graphs}

The definition of the 
MFGs as \textit{directed graphs} opens up 
the application of network-theoretic tools for detecting 
modules of reaction nodes 
and the hierarchical
relationships among them. 
In contrast with methods for undirected graphs, the Markov Stability
framework~\cite{Delvenne2010, Bacik2016} can be used to detect
multi-resolution community structure in directed graphs
(Sec.~\ref{sec:markov}), thus allowing the exploration of the
multiscale organisation of metabolic reaction networks. The modules
so detected reflect subsets of reactions where metabolic fluxes tend 
to be contained.

\begin{figure}[tp]
 \begin{center}
 \includegraphics[width=\textwidth]{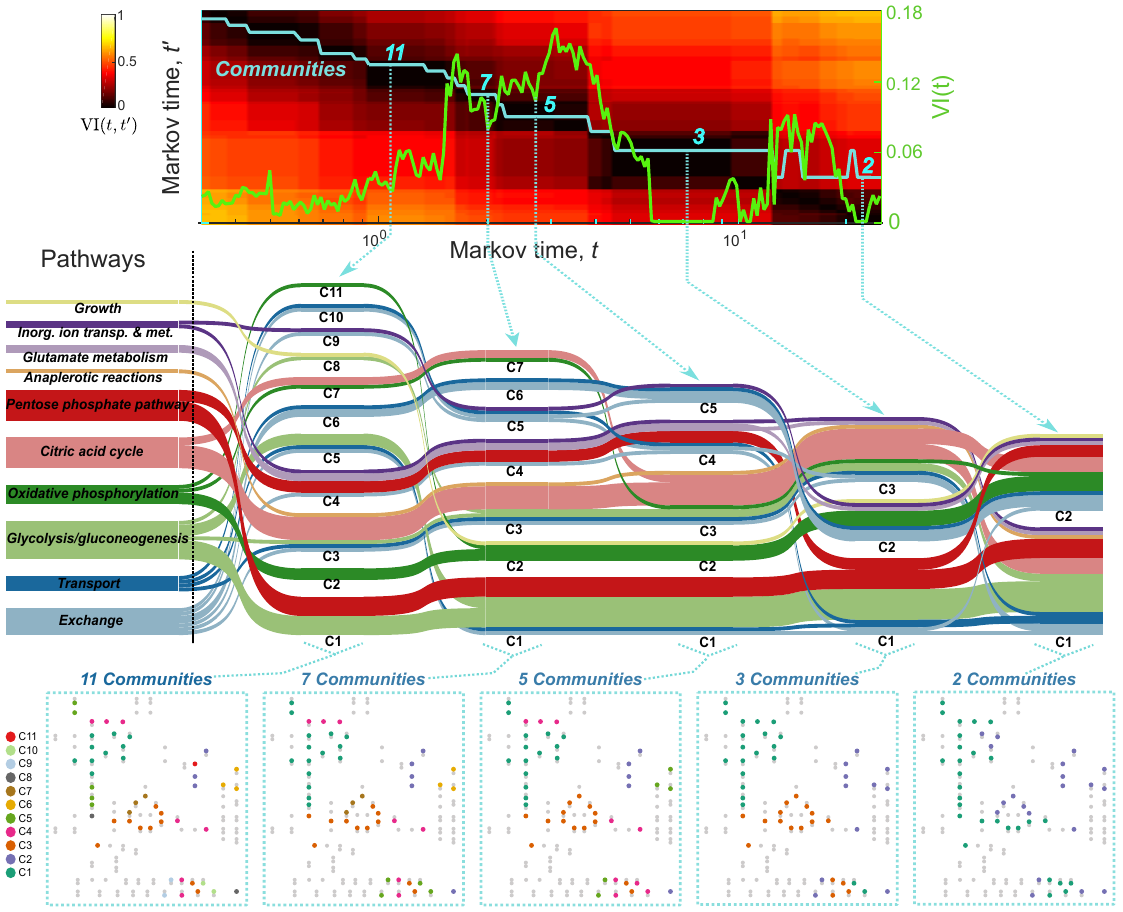}
 \end{center}
 \caption{{\bf Community structure of flux graphs across different scales.}  We applied the Markov Stability method to partition the metabolic flux graph for \emph{E. coli} aerobic growth in glucose ($\M{glc}$) across levels of resolution. The top panel shows the number of communities of the optimal partition (blue line) and two measures of its robustness ($VI(t)$ (green line) and $VI(t,t')$ (colour map)) as a function of the Markov time $t$ (see text and Methods section). The five Markov times selected correspond to robust partitions of the graph into 11, 7, 5, 3, and 2 communities, as signalled by extended low values of $VI(t,t')$ and low values (or pronounced dips) of $VI(t)$.  The Sankey diagram (middle panel) visualises the multiscale organisation of the communities of the flux graph across Markov times, and the relationship of the communities with the biochemical pathways.  The bottom panel shows the five partitions at the selected Markov times. The partition into 3 communities corresponds to that in Figure \ref{fig:networksM}A.}
 \label{fig:VI_alluvialdiagrams}
\end{figure}

Figure~\ref{fig:VI_alluvialdiagrams} illustrates this multiscale
analysis on $\M{glc}$, the MFG of \textit{E. coli} under aerobic
growth in glucose.  By varying the Markov time $t$, a
parameter in the Markov Stability method, we scanned the community
structures at different resolutions.  Our results show that, 
from finer to coarser resolution, the MFG can be partitioned
into 11, 7, 5, 3, and 2 communities of high persistence across
Markov time (extended plateaux over $t$, as shown by
low values of $VI(t,t')$) and high robustness under
optimisation (as shown by dips in $VI(t)$).  For further
details, see Section~\ref{sec:markov} and
Refs.~\cite{Delvenne2010,Schaub2012, Lambiotte2014, Bacik2016}.

The Sankey diagram in Fig.~\ref{fig:VI_alluvialdiagrams}
visualises the pathway composition of the graph partitions and their
relationships across different resolutions.  As we decrease the
resolution (longer Markov times), the reactions in different pathways
assemble and split into different groupings, reflecting both specific
relationships and general organisation principles associated with this
growth condition.  A general observation is that glycolysis is grouped
together with oxidative phosphorylation across most scales,
underlining the fact that those two pathways function as cohesive
metabolic sub-units in aerobic conditions.  In contrast, the exchange
and transport pathways appear spread among multiple partitions across
all resolutions.  This is expected, as exchange/transport are enabling 
functional pathways, in which reactions do not interact amongst 
themselves but rather feed substrates to other pathways.

Other reaction groupings reflect more specific relationships.  For
example, the citric acid cycle (always linked to anaplerotic
reactions) appears as a cohesive unit across most scales, and only
splits in two in the final grouping, reflecting the global
role of the TCA cycle in linking to both glycolysis and oxidative
phosphorylation.  The pentose phosphate pathway, on the other hand, is
split into two groups (one linked to glutamate metabolism and another
one linked to glycolysis) across early scales, only merging into the
same community towards the final groupings. This suggests a more
interconnected flux relationship of the different steps of the
penthose phosphate pathway with the rest of metabolism.
In Figure~SI~2, we present the multiscale
analyses of the reaction communities for the other three 
growth scenarios ($\M{etoh}$, $\M{anaero}$, $\M{lim}$).

\subsection{Using MFGs to analyse hepatocyte metabolism in wild type and PH1 mutant human cells}

To showcase the applicability of our framework to larger metabolic
models, we analyse a model of human hepatocyte (liver) metabolism with
777 metabolites and 2589 reactions~\cite{Pagliarini2016}, which
extends the widely used HepatoNet1 model~\cite{Gille2010} with an additional 50
reactions and 8 metabolites.  This extended model was used in Ref.~\cite{Pagliarini2016}  
to compare wild type cells (WT) and cells affected by the rare disease Primary
Hyperoxaluria Type 1 (PH1), which lack alanine:glyoxylate aminotransferase (AGT) 
due to a genetic mutation. AGT is an enzyme found in peroxisomes  and
its mutation decreases the breakdown of glyoxylate, with subsequent
accumulation of calcium oxalate that leads to liver damage.

\begin{figure}[tp]
 \begin{center}
 \includegraphics[width=\textwidth]{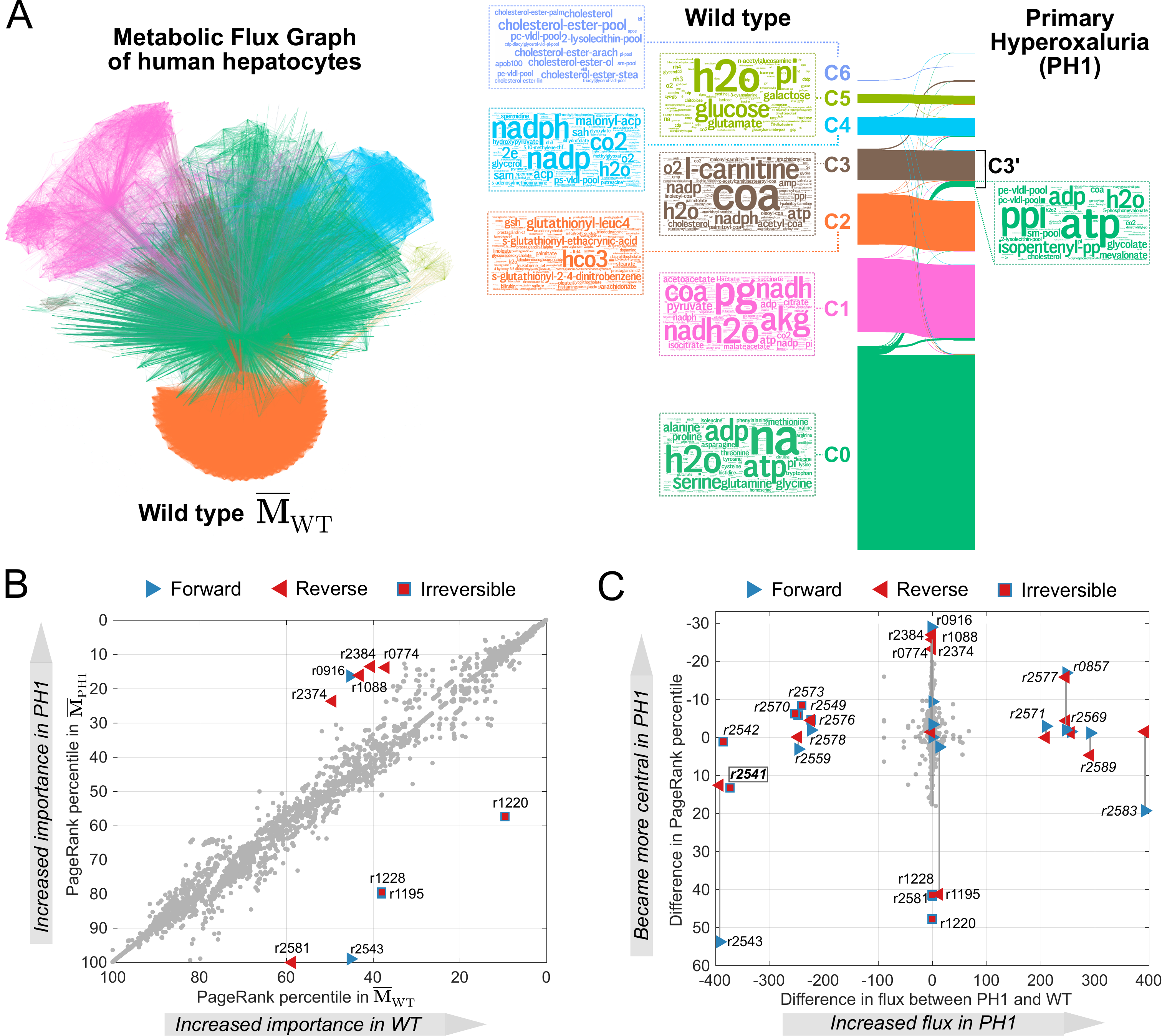}
 \end{center}
 \caption{{\bf MFG analysis of a model of human hepatocyte metabolism and the genetic condition PH1.} (A) Average MFG of wild-type hepatocytes cells over 442 metabolic objectives. The reaction nodes are coloured according to communities in a 7-way partition obtained with Markov Stability.  The Sankey diagramme shows the consistency between the communities in the wild type MFG and the communities independently found in the MFG of the mutated PH1 cells. Word clouds of the most frequent metabolites in the reactions of the WT communities reveal functional groupings (see text).  Under the PH1 mutation, the only large change relates to metabolites that join C3' from community C0 in WT. (B) Comparison of the PageRank percentiles in the WT and PH1 MFGs, with reactions whose rank changes by more than 20 percentiles labelled.  (C) Difference in FBA flux between WT and PH1 vs difference in PageRank percentile between WT and PH1. Reactions whose flux difference is greater than $100\mmol$ (italics) or whose change in PageRank percentile is greater than 20 are labelled. The differences in centrality (PageRank) provide complementary information, revealing additional important reactions affected by the PH1 mutation that knocks out reaction r2541.}
 \label{fig:hepatocytes}
\end{figure}

Following~\cite{Pagliarini2016}, we first obtain 442 FBA solutions for
different sets of metabolic objectives for both the wild type (WT)
model and the PH1 model lacking AGT (reaction r2541).
We then generate the corresponding 442 MFGs for each WT and
PH1, and obtain the averages over each ensemble:
$\overline{\mbf{M}}_{\mathrm{WT}}$ and
$\overline{\mbf{M}}_{\mathrm{PH1}}$.  Of the 2589 reactions in the
model, 2448 forward and 1362 reverse reactions are present in at least
one of the FBA solutions.  Hence the average MFGs have 3810 nodes
each (see Supplementary Spreadsheet for full details about the
reactions).
%

Figure~\ref{fig:hepatocytes}A shows the MFG for the wild type
($\overline{\mbf{M}}_{\mathrm{WT}}$) coloured according to a robust
partition into 7 communities obtained with Markov Stability. The seven
communities are broadly linked to amino acid metabolism (C0), energy
metabolism (C1 and C5), glutathione metabolism (C2), fatty acid and
bile acid metabolism (C3 and C4) and cholesterol metabolism and lipoprotein
particle assembly (C6).  As expected, the network community structure
of the MFG is largely preserved under the AGT mutation: the Sankey
diagramme in Fig.~\ref{fig:hepatocytes}B shows a remarkable match
between the partitions of $\overline{\mbf{M}}_{\mathrm{WT}}$ and
$\overline{\mbf{M}}_{\mathrm{PH1}}$ found independently with Markov
Stability. Despite this similarity, our method also identified subtle
but important differences between the healthy and diseased networks.
In particular, C3' in $\overline{\mbf{M}}_{\mathrm{PH1}}$ receives 60
reactions, almost all taking place in the peroxisome and linked to
mevalonate and iso-pentenyl pathways, as well as highly central
transfer reactions of PP$_i$, O$_2$ and H$_2$O$_2$ 
between the peroxisome and the cytosol (r1152, r0857, r2577) .

Overall, the centrality (PageRank) of most reactions in the MFG is relatively
unaffected by the PH1 mutation, as shown by the
good correlation between the PageRank percentiles in
$\overline{\mbf{M}}_{\mathrm{WT}}$ and
$\overline{\mbf{M}}_{\mathrm{PH1}}$ in Figure~\ref{fig:hepatocytes}C. 
Yet, there are notable exceptions, and the reactions that exhibit the largest change in
PageRank centrality (labelled in Fig.~\ref{fig:hepatocytes}B) provide
biological insights into the disease state.
%
%
%
Specifically, the four reactions (r0916, r1088, r2384, r2374) 
that undergo the largest increase in centrality from $\overline{\mbf{M}}_{\mathrm{WT}}$ 
to $\overline{\mbf{M}}_{\mathrm{PH1}}$  
are related to the transfer of citrate out of the cytosol in exchange for oxalate and PEP; 
whereas those with the largest decrease of PageRank  from $\overline{\mbf{M}}_{\mathrm{WT}}$ 
to $\overline{\mbf{M}}_{\mathrm{PH1}}$ are related to VLDL-pool reactions (r1228, r1195, r1220) and to transfers of hydroxypyruvate and alanine from peroxisome to cytosol (r2581, r2543).
It is worth remarking that although oxalate and citrate reactions are directly linked to
metabolic changes associated with the PH1 diseased state, 
none of them exhibits large changes in their flux predicted by FBA, 
yet they show large changes in PageRank centrality.

These observations underscore how the information provided by our
network analysis provides complementary information to the analysis of 
FBA fluxes alone. As shown in Figure~\ref{fig:hepatocytes}D, there is a group of
reactions (labelled with italics in the Figure) that exhibit large gains or decreases 
in their flux under the PH1 mutation, yet they only undergo relatively small changes 
in their PageRank scores. Closer inspection reveals that most of these reactions are
close to the AGT reaction (r2541, highlighted in the Figure) in the pathway and involve
the conversion of glycolate, pyruvate, glycine, alanine and
serine. Hence the changes in flux follow from the \textit{local
rearrangement} of flows as a consequence of the deletion of
reaction r2541.  On the other hand, the citrate and oxalate reactions (r0916, r1088, r2384, r2374)
with large changes in their centrality yet undergo small changes in flux, thus reflecting 
\textit{global changes} in the flux structure of the network. 
Importantly, the transport reactions of
O$_2$, H$_2$O$_2$, serine and hydroxypyruvate between cytosol and
peroxisome (r0857, r2577, r2583, r2543) all undergo large changes both in centrality and flux,
highlighting the importance of peroxisome transfer reactions in PH1.  
We provide a full spreadsheet with these analyses as Supplementary
Material for the interested reader.

\section{Discussion}

Metabolism is commonly understood in terms of functional
pathways interconnected into metabolic networks~\cite{King2015}, i.e.,
metabolites linked by arrows representing enzymatic reactions between
them as in Figure~\ref{fig:Model_A_Dnorm}A.  
However, such standard representations are not
amenable to rigorous graph-theoretic analysis. 
Fundamentally different graphs can be constructed from such
metabolic reactions depending on the chosen representation
of species/interactions as nodes/edges, e.g., reactions as nodes;
metabolites as nodes; or both reaction and metabolites as
nodes~\cite{Palsson2006}.  Each of those graphs can be directed or
undirected and with weighted links computed according to different
rules.  The choices and subtleties in graph construction are crucial
both to capture the relevant metabolic information and to interpret
their topological properties~\cite{Ouzounis2000,Arita2004}.

Here, we have presented a flux-based strategy to build graphs for
metabolic networks.  Our graphs have reactions as nodes and directed weighted
edges representing the flux of metabolites produced by a source
reaction and consumed by a target reaction.  This principle is
applied to build both `blueprint' graphs (PFG), which summarise
probabilistically the fluxes of the whole metabolism of an organism, 
as well as context-specific graphs (MFGs), which reflect specific
environmental conditions.  The blueprint Probabilistic Flux Graph has
edge weights equal to the probability that source/target reactions
produce/consume a molecule of a metabolite chosen at random in the
absence of any other information, and can thus be used 
when the stoichiometric matrix is the only information available.
The PFG construction naturally tames the
over-representation of pool metabolites without the need to remove
them from the graph arbitrarily, as often done in the
literature~\cite{Ma2003,Silva2007,Samal2011,Kreimer2008}. 
Context-specific Metabolic Flux Graphs (MFGs) can incorporate the effect of the environment, e.g.,
with edge weights corresponding to the total flux of metabolites between reactions
as computed by Flux Balance Analysis (FBA). FBA solutions
for different environments can then be used to build different metabolic graphs
in different growth conditions.

The proposed graph constructions provide complementary tools for studying the
organisation of metabolism and can be embedded into any
FBA-based modelling pipeline. Specifically, the PFG relies on the
availability of a well-curated stoichiometric matrix, which is
produced with metabolic reconstruction techniques that typically
precede the application of FBA. The MFG, on the other hand, explicitly
uses the FBA solutions in its construction. Both methods provide a
systematic framework to convert genome-scale metabolic models into a
directed graph on which analysis tools from network theory can be applied.

To exemplify our approach, we built and analysed PFG and MFGs for the
core metabolism of \textit{E. coli}.  Through the analysis of
topological properties and community structure of these graphs, we
highlighted the importance of weighted directionality in metabolic
graph construction, and revealed the flux-mediated relationships
between functional pathways under different environments.  In
particular, the MFGs capture specific metabolic adaptations such as
the glycolytic-gluconeogenic switch, overflow metabolism, and the
effects of anoxia.  The proposed graph construction can be readily 
applied to large genome-scale metabolic
networks~\cite{Ravasz2002, Smart2008, Ma2004, Samal2006, folch2015}.
  
To illustrate the scalability of our analyses to larger metabolic
models, we studied a genome-scale model of a large metabolic model of
human hepatocytes with around 3000 reactions in which we compared the
wild type and a mutated state associated with the disease PH1 under more than
400 metabolic conditions~\cite{Pagliarini2016}.  Our network analysis
of the MFGs revealed a consistent organisation of the
reaction graph, which is highly preserved under the mutation.
Our analysis also identified notable changes in the network centrality score 
and community structure of certain reactions, which is linked to key biological processes
in PH1. Importantly, network measures computed from
the MFGs reveal complementary information to that provided by the sole
analysis of perturbed FBA fluxes.

Our flux graphs provide a systematic connection between network theory
and constraint-based methods widely employed in metabolic
modelling~\cite{Orth2010a,Rabinowitz2012, Samal2006,Smart2008}, thus
opening avenues towards environment-dependent, graph-based analyses of
cell metabolism.  An area of interest for future research is the use of MFGs 
to study how network measures of flux graphs can help
characterise metabolic conditions that maximise the efficacy of drug
treatments or disease-related distortions, e.g., cancer-related
metabolic signatures~\cite{Csermely2005,Chang2010,Folger2011,Vaitheesvaran2015}.
In particular, MFGs can quantify metabolic robustness via graph
statistics upon removal of reaction nodes~\cite{Smart2008}.

The proposed graph construction framework can be extended in different
directions.  The core idea behind our framework is the distinction
between production and consumption fluxes, and how to encode both in
the links of a graph. This general principle can also be used to build
other potentially useful graphs. For example, two other graphs that
describe relationships between reactions are:
\begin{align}
  \textbf{Competition flux graph:} \quad \quad \nDc & = \frac{1}{n} \,
  \mbf{S}_{2m}^{- \, T} \lp \mbf{W}_{-}^{\dagger}
  \rp^2 {\mbf{S}_{2m}^{-}}
 \label{eq:Dc} \\
  \textbf{Synergy flux graph:} \quad \quad \nDs & = \frac{1}{n} \,
  \mbf{S}_{2m}^{+ \, T} \lp 
  \mbf{W}_{+}^{\dagger} \rp^2 {\mbf{S}_{2m}^{+}} .
  \label{eq:Ds}
\end{align}
The competition and synergy graphs are undirected and their edge
weights represent the probability that two reactions consume ($\nDc$)
or produce ($\nDs$) metabolites chosen at random. The
corresponding FBA versions of \textit{competition and synergy flux graphs},
which follow directly from~\eqref{eq:Mv},~\eqref{eq:Dc} and
\eqref{eq:Ds} could help reveal further relationships
between metabolic reactions in the cell. These graphs will be the subject of
future studies.

Our approach could also be extended to include dynamic adaptations of
metabolic activity: by using dynamic extensions of
FBA~\cite{Waldherr2015,Rugen2015}; by incorporating
static~\cite{Colijn2009} or time-varying~\cite{Oyarzun2011b} enzyme
concentrations; or by considering kinetic models (with kinetic constants when
available) to generate probabilistic reactions fluxes in the sense of stochastic chemical kinetics~\cite{gillespie1977,Oyarzun2015}.  
Of particular interest to metabolic modelling, we envision that MFGs could provide a
novel route to evaluate the robustness of FBA
solutions~\cite{Gudmundsson2010, Orth2010a} by exploiting the
non-uniqueness of the MFG from each FBA solution in the space of
graphs.  Such results could enhance the interface between network
science and metabolic analysis, allowing for the systematic
exploration of the system-level organisation of metabolism in response
to environmental constraints and disease states.

\appendix
\section{Methods}

\subsection{Flux balance analysis}
\label{sec:fba}

Flux Balance Analysis (FBA)~\cite{Orth2010a, Rabinowitz2012} is a
widely-adopted approach to analyse metabolism and cellular growth. FBA
calculates the reaction fluxes that optimise growth in specific
biological contexts.  The main hypothesis behind FBA is that cells
adapt their metabolism to maximise growth in different biological
conditions. The conditions are encoded as constraints on the fluxes of
certain reactions; for example, constraints reactions that import
nutrients and other necessary compounds from the exterior.

The mathematical formulation of the FBA is described in the following
constrained optimisation problem:
\begin{equation}
  \begin{split}
    \text{maximise:}& \quad \mbf{c}^{T}\mbf{v} \\
    \text{subject to} \,\,  & \left\{
    \begin{array}{c}
      \mbf{Sv} = 0 \\
      \mbf{v}_{\mathrm{lb}} \leq \mbf{v} \leq \mbf{v}_{\mathrm{ub}},
    \end{array}
    \right.
  \end{split}
  \label{eq:fba}
\end{equation}
where $\mbf{S}$ is the stoichiometry matrix of the model, $\mbf{v}$
the vector of fluxes, $\mbf{c}$ is an indicator vector (\ie $c(i)=1$
when $i$ is the biomass reaction and zero everywhere else) so that
$\mbf{c}^{T} \mbf{v}$ is the flux of the biomass reaction.  The
constraint $\mbf{S}\mbf{v} = 0$ enforces mass-conservation at
stationarity, and $\mbf{v}_{\mathrm{lb}}$ and $\mbf{v}_{\mathrm{ub}}$
are the lower and upper bounds of each reaction's flux.  Through these
vectors, one can encode a variety of different
scenarios~\cite{Orth2010}. The biomass reaction represents the most
widely-used flux that is optimised, although there are others can be
used as well~\cite{Feist2010,Schuetz2007}.

In our simulations, we set the individual carbon intake rate to
18.5~$\mmol$ for every source available in each scenario. We allowed
oxygen intake to reach the maximum needed in to consume all the carbon
except in the anaerobic condition scenario, in which the upper bound
for oxygen intake was $0~\mmol$. In the scenario with limited phosphate
and ammonium intake, the levels of NH$_4$ and phosphate intake were
fixed at $4.5~\mmol$ and $3.04~\mmol$ respectively (a reduction of
50\% compared to a glucose-fed aerobic scenario with no restrictions).

\subsection{Markov Stability community detection framework}
\label{sec:markov}

We extract the communities in each network using the Markov Stability
community detection framework~\cite{Delvenne2010,Delvenne2013}. This
framework uses diffusion processes on the network to find groups of
nodes (i.e., communities) that retain flows for longer than one would
expect on a comparable random network; in addition, Markov Stability
incorporates directed flows seamlessly into the
analysis~\cite{Lambiotte2014, Beguerisse2014}.

The diffusion process we use is a continuous-time Markov process on
the network.  From the adjacency matrix $\mbf{G}$ of the graph (in our case, the RAG, PFG or MFG), 
we construct a rate matrix for the process: $\mbf{M} = \mbf{K}_{\mathrm{out}}^{-1}{\mbf{G}}$, where
$\mbf{K}_{\mathrm{out}}$ is the diagonal matrix of out-strengths,
$k_{\mathrm{out},i}=\sum_{j}g_{i,j}$. When a node has no outgoing
edges then we simply let $k_{\mathrm{out},i}=1$.  In general, a
directed network will not be strongly-connected and thus a Markov
process on $\mbf{M}$ will not have a unique steady state.  To ensure
the uniqueness of the steady state we must add a {\it teleportation}
component to the dynamics by which a random walker visiting a node can
follow an outgoing edge with probability $\lambda$ or jump (teleport)
uniformly to any other node in the network with probability
$1-\lambda$~\cite{Page1999}. The rate matrix of a Markov process with
teleportation is:
\begin{equation}
  \mbf{B} = \lambda \mbf{M} + \frac{1}{N}\left[(1-\lambda)
    \mbf{I}_N + \lambda \,
    \mathrm{diag}(\mbf{a})\right]{\mathbf{1}\mathbf{1}^T},
\end{equation}
where the $N \times 1$ vector $\mbf{a}$ is an indicator for dangling
nodes: if node $i$ has no outgoing edges then $a_i=1$, and $a_i=0$
otherwise. Here we use $\lambda=0.85$.  The Markov process is
described by the ODE:
\begin{equation}
  \dot{\mbf{x}} = -\mbf{L}^T\mbf{x},
  \label{eq:kolmo}
\end{equation}
where $\mbf{L} = \mbf{I}_N - \mbf{B}$.  The solution of
\eqref{eq:kolmo} is $\mbf{x}(t) = e^{-t\mbf{L}^T}\mbf{x}(0)$ and its
stationary state (i.e., $\dot{\mbf{x}}=0$) is
$\mbf{x}=\boldsymbol{\pi}$, where $\boldsymbol{\pi}$ is the leading
left eigenvector of $\mbf{B}$.

A hard partition of the graph into $C$ communities can be encoded into
the $N\times C$ matrix $\mbf{H}$, where $h_{ic}=1$ if node $i$ belongs
to community $c$ and zero otherwise. The $C\times C$ {\it clustered
  autocovariance matrix} of~\eqref{eq:kolmo}~is
\begin{equation}
  \mbf{R}(t, \mbf{H}) = \mbf{H}^T\left(\boldsymbol{\Pi}e^{-t\mbf{L}^T}
  - \boldsymbol{\pi}\boldsymbol{\pi}^T\right)\mbf{H},
\end{equation}
and the entry $(c,s)$ of $\mbf{R}(t,\mbf{H})$ measures how likely it
is that a random walker that started the process in community $c$
finds itself in community $s$ after time $t$ when at stationarity. The
diagonal elements of $\mbf{R}(t,\mbf{H})$ thus record how good the
communities in $\mbf{H}$ are at retaining flows. The {\it Markov
  stability of the partition} is then defined as
\begin{equation}
  r(t, \mbf{H}) = \trace \, \mbf{R}(t, \mbf{H}).
  \label{eq:stab}
\end{equation}
The optimised communities are obtained by maximising the cost
function~\eqref{eq:stab} over the space of all partitions for every
time~$t$ to obtain an optimised partition $\widehat{\mathcal{P}}(t)$.  
This optimisation is NP-hard; hence with no guarantees of
optimality.  Here we use the Louvain greedy optimisation
heuristic~\cite{Blondel2008}, which is known to give high quality
solutions $\widehat{\mathcal{P}}(t)$ in an efficient manner.  
The value of the Markov time $t$, i.e. the duration of the Markov process, can be understood as a
resolution parameter for the partition into
communities~\cite{Delvenne2010, Schaub2012}. In the limit $t\to 0$,
Markov stability will assign each node to its own community; as $t$
grows, we obtain larger communities because the random walkers have
more time to explore the network~\cite{Delvenne2013}.  We scan through
a range of values of $t$ to explore the multiscale community structure
of the network.  The code for Markov Stability can be found at
\url{github.com/michaelschaub/PartitionStability}.

To identify the important partitions across time, we use two criteria
of robustness~\cite{Schaub2012}.  Firstly, we optimise \eqref{eq:stab}
100 times for each value of $t$ and we assess the consistency of the
solutions found.  A relevant partition should be a robust outcome of
the optimisation, i.e., the ensemble of optimised solutions should be
similar as measured with the normalised variation of
information~\cite{Meila2007}:
\begin{equation}
\label{eq:VI}
 \textit{VI} (\mathcal P, \mathcal P') = \dfrac{2 \Omega(\mathcal P,
   \mathcal P') - \Omega(\mathcal P) - \Omega(\mathcal P')}{\log(n)},
\end{equation}
where $\Omega(\mathcal P) = -\sum_{\mathcal C} p(\mathcal C) \log
p(\mathcal C)$ is a Shannon entropy and $p(\mathcal C)$ is the
relative frequency of finding a node in community $\mathcal C$ in
partition $\mathcal P$.  We then compute the average variation of
information of the ensemble of solutions from the $\ell =100$ Louvain
optimisations $\mathcal{P}_i(t)$ at each Markov time $t$:
\begin{align}
   \text{\VIt} = \dfrac{1}{\ell (\ell-1)} \sum_{i\neq j} \VI
   (\mathcal{P}_i(t),\mathcal{P}_j(t)).
\end{align}
If all Louvain runs return similar partitions, then \VIt is small,
indicating robustness of the partition to the optimisation. Hence we
select partitions with low values (or dips) of \VIt.  Secondly,
relevant partitions should also be optimal across Markov time, as
indicated by a low values of the cross-time variation of information:
\begin{align}
    \text{\VItt} = \VI
    (\widehat{\mathcal{P}}(t),\widehat{\mathcal{P}}(t')).
\end{align}
Therefore, we also search for partitions with extended low value plateaux of \VItt~\cite{Lambiotte2014,Schaub2012,Bacik2016}.

\section*{Acknowledgments}

M.B.D.~acknowledges support from the James S. McDonnell Foundation
Postdoctoral Program in Complexity~Science/Complex~Systems Fellowship
Award (\#220020349-CS/PD Fellow), and the Oxford-Emirates Data Science
Lab.  G.B.~acknowledges the support from the Spanish Ministry of
Economy FPI Program (BES-2012-053772). D.O.~acknowledges support from
an Imperial College Research Fellowship and from the Human Frontier
Science Program through a Young Investigator Grant
(RGY0076-2015). J.P.~acknowledges the support from the Spanish
Ministry of Economy and EU FEDER through the SynBioFactory project
(CICYT DPI2014-55276-C5-1). M.B.~acknowledges funding from the EPSRC
through grants EP/I017267/1 and EP/N014529/1.

\paragraph*{Data statement:}
No new data was generated during the course of this research.

\newpage
\appendix

\setcounter{figure}{0}
\setcounter{section}{0}
\setcounter{equation}{0}
\renewcommand\thesection{S\arabic{section}}
\renewcommand\thefigure{S\arabic{figure}}
\renewcommand\theequation{S\arabic{equation}}

\centerline{\Large {\bf Supplementary Information}}

\section{Relation of the PFG with a directed version of the RAG}
\label{sec:Adir_vs_D_SI}
A directed version of the RAG~\eqref{eq:A} could in principle be
obtained from the boolean production/consumption matrices
$\widehat{\mbf{S}}_{2m}^{+}$ and $\widehat{\mbf{S}}_{2m}^{-}$ as
follows.  Projecting onto the space of reactions gives the
$2m\times2m$ (asymmetric) adjacency matrix
\begin{equation}
  \nD = {\widehat{\mbf{S}}_{2m}^{+\,T}}\,\widehat{\mbf{S}}_{2m}^{-},
  \label{eq:D}
\end{equation} 
where the entries $D_{ij}$ represent the total number of metabolites
{\it produced} by reaction $R_{i}$ that are {\it consumed} by reaction
$R_{j}$.  A directed version of the Reaction Adjacency Graph on $m$
nodes (directly comparable to the standard RAG) is then
\begin{align} 
  \Adir &= 
  \begin{bmatrix}\mbf{I}_m & \mbf{I}_m \end{bmatrix} 
  \nD
  \begin{bmatrix}\mbf{I}_m \\ \mbf{I}_m \end{bmatrix}.
  \label{eq:A-D}
\end{align}
Clearly, when the metabolic model contains only reversible reactions,
(\ie the reversibility vector is all ones, $\mbf{r}=
\mathbbm{1}_{m}$), it follows that $\Adir={\nA}$.

Although $\Adir$ does not include spurious edges introduced by
non-existent backward reactions, its structure is still obscured by
the effect of uninformative connections created by pool metabolites.

\section{Details of the toy metabolic network}
\label{sec:toy_model}

As an illustration of the graph construction, the toy metabolic
network in Fig.~\ref{fig:ToyModel_networks} was taken from
Ref.~\cite{Rabinowitz2012}. The graph matrices for this model are as
follows:
\begin{itemize}
\item Reaction Adjacency Graph, Eq.~\eqref{eq:A}:
{\small
\begin{equation*}
\mathbf{A} = 
\widehat{\mathbf{S}}^T\widehat{\mathbf{S}} = 
\left[ \begin{array}{rrrrrrrr} 
    1 & 1 & 0 & 0 & 0 & 0 & 0 & 0 \\
    1 & 2 & 1 & 1 & 0 & 0 & 0 & 0 \\
    0 & 1 & 2 & 1 & 1 & 0 & 0 & 1 \\
    0 & 1 & 1 & 2 & 0 & 1 & 1 & 1 \\ 
    0 & 0 & 1 & 0 & 2 & 1 & 0 & 2 \\
    0 & 0 & 0 & 1 & 1 & 2 & 1 & 2 \\
    0 & 0 & 0 & 1 & 0 & 1 & 1 & 1 \\
    0 & 0 & 1 & 1 & 2 & 2 & 1 & 3 \\ 
  \end{array}
\right].
\end{equation*}
}
\item Probabilistic Flux Graph, Eq.~\eqref{eq:Dp}:
{\small
\begin{equation*} \nDp = 
  \frac{1}{n} \, \mbf{S}_{2m}^{+ \, T}   \lp \mbf{W}_{+}^{\dagger}
  \mbf{W}_{-}^{\dagger} \rp  {\mbf{S}_{2m}^{-}} = 
\left[
  \begin{array}{rrrrrrrrrr}  
  & R_1 & R_2 & R_3 & R_4 & R_5 & R_6 & R_7 & R_8 & R_{4r}\\
  R_1 & 0 & 0.2 & 0 & 0 & 0 & 0 & 0 & 0 & 0\\
  R_2 & 0 & 0 & 0.05 & 0.05 & 0 & 0 & 0 & 0 & 0\\
  R_3 & 0 & 0 & 0 & 0 & 0.1 & 0 & 0 & 0.1 & 0\\
  R_4 & 0 & 0 & 0 & 0 & 0 & 0.04 & 0.04 & 0.08 & 0.04\\
  R_5 & 0 & 0 & 0 & 0 & 0 & 0 & 0 & 0.1 & 0\\
  R_6 & 0 & 0 & 0 & 0 & 0 & 0 & 0 & 0.1 & 0\\
  R_7 & 0 & 0 & 0 & 0 & 0 & 0 & 0 & 0 & 0\\
  R_8 & 0 & 0 & 0 & 0 & 0 & 0 & 0 & 0 & 0\\
  R_{4r} & 0 & 0 & 0.05 & 0.05 & 0 & 0 & 0 & 0 & 0 \\
\end{array}
  \right].
\end{equation*}
}
\item Metabolic Flux Graph for FBA scenario 1, Eq.~\eqref{eq:Mv}:
{\small
\begin{equation*}
\begin{array}{crrr}  
  & \mathbf{v}_{\mathrm{lb}_1} & \mathbf{v}_{\mathrm{ub}_1} & \mathbf{v}_1^* \\ 
  R_1: & 10 & 10 & 10 \\ 
  R_2: & 0 & 10 & 10 \\ 
  R_3: & 0 & 10 & 4.992 \\ 
  R_4: & -10 & 10 & 5.008 \\ 
  R_5: & 0 & 10 & 2.492 \\ 
  R_6: & 0 & 10 & 0.008 \\ 
  R_7: & 0 & 10 & 0 \\ 
  R_8: & 0 & 10 & 2.5 \\ 
  R_{4r}: & -10 & 10 & 0
\end{array} \qquad
 \nM{1} = \left[
\begin{array}{rrrrrrrr}                                             
 & R_1 & R_2 & R_3 & R_4 & R_5 & R_6 & R_8 \\   
R_1 & 0 & 10 & 0 & 0 & 0 & 0 &  0 \\
R_2 & 0 & 0 & 4.992 & 5.008 & 0 & 0 & 0 \\ 
R_3 & 0 & 0 & 0 & 0 & 2.492 & 0 & 2.5 \\ 
R_4 & 0 & 0 & 0 & 0 & 0 & 0.008 & 5 \\ 
R_5 & 0 & 0 & 0 & 0 & 0 & 0 & 2.492 \\ 
R_6 & 0 & 0 & 0 & 0 & 0 & 0 & 0.008 \\ 
R_8 & 0 & 0 & 0 & 0 & 0 & 0 & 0 \\ 
\end{array}
  \right].
\end{equation*}
}

\item Metabolic Flux Graph for FBA scenario 2, , Eq.~\eqref{eq:Mv}:
{\small
\begin{equation*}
\begin{array}{crrr}
  & \mathbf{v}_{\mathrm{lb}_2} & \mathbf{v}_{\mathrm{ub}_2} & \mathbf{v}_2^* \\
  R_1: & 10 & 10 & 10 \\    
  R_2: & 0 & 10 & 10 \\     
  R_3: & 0 & 10 & 3.877 \\      
  R_4: & -10 & 10 & 6.123 \\    
  R_5: & 0 & 10 & 1.877 \\      
  R_6: & 0 & 10 & 0.123 \\      
  R_7: & 2 & 10 & 2 \\      
  R_8: & 0 & 10 & 2 \\      
  R_{4r}: & -10 & 10 & 0
\end{array} \qquad 
\nM{2} = \left[
\begin{array}{rrrrrrrrr}                                             
 & R_1 & R_2 & R_3 & R_4 & R_5 & R_6 & R_7 & R_8 \\                    
R_1 & 0 & 10 & 0 & 0 & 0 & 0 & 0 & 0 \\
R_2 & 0 & 0 & 3.877 & 6.123 & 0 & 0 & 0 & 0 \\ 
R_3 & 0 & 0 & 0 & 0 & 1.877 & 0 & 0 & 2 \\ 
R_4 & 0 & 0 & 0 & 0 & 0 & 0.123 & 2 & 4 \\ 
R_5 & 0 & 0 & 0 & 0 & 0 & 0 & 0 & 1.877 \\ 
R_6 & 0 & 0 & 0 & 0 & 0 & 0 & 0 & 0.123 \\ 
R_7 & 0 & 0 & 0 & 0 & 0 & 0 & 0 & 0 \\ 
R_8 & 0 & 0 & 0 & 0 & 0 & 0 & 0 & 0 \\ 
\end{array}    
  \right].
\end{equation*}
}
\end{itemize}

\section{ Reaction communities in context-free graphs of the core \textit{E.~coli} metabolic model}
\label{sec:AvsD_comms_SI}

\subsection{Reaction Adjacency Graph, ${\nA}$}

\begin{figure}[tp]
 \begin{center}
   \includegraphics[width=\textwidth]{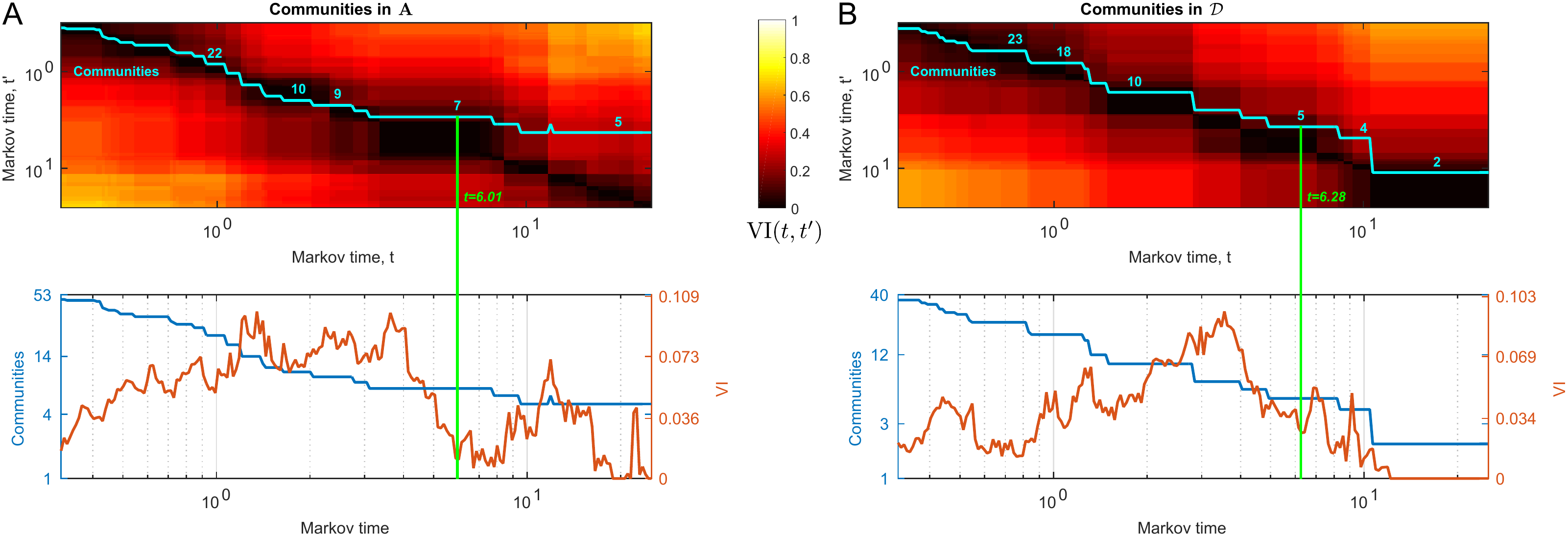}
 \end{center}
 \caption{{\bf Community structure in the template networks $\nA$ and
     $\nDp$.} (A) Communities in $\nA$. Top plot: Variation of
   Information (VI) of the best partition found at Markov time $t$
   with every other partition at time $t'$. Bottom plot: Number of
   communities and VI of the ensemble of solutions found at each
   Markov time. A robust partition into seven communities is found at
   $t=6.01$. (B) Communities and VI in $\nDp$. A robust partition into
   five communities is found at $t=6.28$.}
 \label{fig:Markov_Stability_template_networks}
\end{figure}

A robust partition into seven communities in the RAG was found at
Markov time $t=6.01$
(Fig.~\ref{fig:Markov_Stability_template_networks}A). The communities
at this resolution (Fig.~\ref{fig:AvsD_comms}E) are:

\begin{itemize}
\item Community \C{1}{\nA} contains all the reactions that consume or
  produce ATP and water (two pool metabolites). Production of ATP
  comes mostly from oxidative phosphorylation (ATPS4r) and substrate
  level phosphorylation reactions such as phosphofructokinase (PFK),
  phosphoglicerate kinase (PGK) and succinil-CoA synthase
  (SUCOAS). Reactions that consume ATP include glutamine synthetase
  (GLNS) and ATP maintenance equivalent reaction (ATPM).  The
  reactions L-glutamine transport via ABC system (GLNabc), acetate
  transport in the form of phosphotransacetilase (PTAr), and acetate
  kinase (ACKr) are also part of this community.  Additionally,
  \C{1}{\nA} (green) contains also reactions that involve
  H$_2$O. Under normal conditions water is assumed to be abundant in
  the cell, thus the biological link that groups these reactions
  together is tenuous.
  
\item Community \C{2}{\nA} includes the reactions NADH dehydrogenase
  (NADH16), cytochrome oxidase (CYTBD), and transport and exchange
  reactions. These two reactions involve pool metabolites (such as
  H$^+$) which create a large number of connection.  Other members
  include fumarate reductase (FR7) and succinate dehydrogenase (SUCDi)
  which couple the TCA cycle with the electron transport chain
  (through ubiquinone-8 reduction and ubiquinol-8
  oxidation). Reactions that include export and transport of most
  secondary carbon sources (such as pyruvate, ethanol, lactate,
  acetate, malate, fumarate, succinate or glutamate) are included in
  the community as well. These reactions are included in the community
  because of their influence in the proton balance of the cell. Most
  of these reactions do not occur under normal circumstances. This
  community highlights the fact that in the absence of biological
  context, many reactions that do not normally interact can be grouped
  together.
  
\item Community \C{3}{\nA} contains reactions that produce or
  consume nicotinamide adenine dinucleotide (NAD$^+$), nicotinamide
  adenine dinucleotide phosphate (NADP$^{+}$), or their reduced
  variants NADH and NADPH. The main two reactions of the community are
  NAD(P) transhydrogenase (THD2) and NAD$^+$ transhydrogenase
  (NADTRHD). There are also reactions related to the production of
  NADH or NADPH in the TCA cycle such as isocitrate dehydrogenase
  (ICDHyr), 2-oxoglutarate dehydrogenase (AKGDH) and malate
  dehydrogenase (MDH).  The community also includes reactions that are
  not frequently active such as malic enzime NAD (ME1) and malic
  enzime NADH (ME2) or acetate dehydrogenase (ACALD) and ethanol
  dehydrogenase (ALCD2x).

\item Community \C{4}{\nA} contains the main carbon intake of
  the cell (glucose), the initial steps of glycolysis, and most of the
  pentose phosphate shunt.  These reactions are found in this
  community because the metabolites involved in these reactions (e.g.,
  alpha-D-ribose-5-phosphate (r5p) or D-erythrose-4-phosphate (e4p))
  are only found in these reactions. This community includes the
  biomass reaction due to the number of connections created by growth
  precursors.
  
\item Communities \C{5}{\nA}, \C{6}{\nA} and \C{7}{\nA}
 are small communities that contain oxygen intake, ammonium intake and 
 acetaldehyde secretion reactions, respectively.
\end{itemize}

\subsection{Probabilistic Flux Graph, $\nDp$}
A robust partition into five communities in the PFG was found at
Markov time $t=6.28$
(Fig.~\ref{fig:Markov_Stability_template_networks}B). The communities
at this resolution (Fig.~\ref{fig:AvsD_comms}C) are:

\begin{itemize}
\item Community \C{1}{\nDp} includes the first half of the
  glycolysis and the complete pentose phosphate pathway.  The
  metabolites that create the connections among these reactions such
  as D-fructose, D-glucose, or D-ribulose.
  
\item Community \C{2}{\nDp} contains the main reaction that
  produces ATP through substrate level (PGK, PYK, ACKr) 
  and oxidative phosphorylation (ATPS4r). The flow of
  metabolites among the reactions in this community includes some pool
  metabolites such as ATP, ADP, H$_2$0, and phosphate. However, there
  are connections created by metabolites that only appear in a handful
  of reactions such as adenosine monophosphate (AMP) whose sole
  producer is phosphoenolpyruvate synthase (PPS) and its sole consumer
  is ATPS4r. This community also contains the biomass reaction.
  
\item Community \C{3}{\nDp} includes the core of the citric
  acid (TCA) cycle such as citrate synthase (CS), aconitase A/B
  (ACONTa/b), and anaplerotic reactions such as malate synthase
  (MALS), malic enzyme NAD (ME1), and malic enzyme NADP (ME2). This
  community also includes the intake of cofactors such as CO$_{2}$.
  
\item Community \C{4}{\nDp} contains reactions that are secondary
  sources of carbon such as malate and succinate, as well as oxidative
  phosphorilation reactions.
  
\item Community \C{5}{\nDp} contains some reactions part of the
  pyruvate metabolism subsystem such as D-lactate dehydrogenase
  (LDH-D), pyruvate formate lyase (PFL) or acetaldehyde dehydrogenase
  (ACALD). In addition, it also includes the tranport reaction for the
  most common secondary carbon metabolites such as lactate, formate,
  acetaldehyde and ethanol.
\end{itemize}

\section{Reaction communities in Metabolic Flux Graphs of \textit{E.~coli} metabolism
under different biological scenarios}
\label{sec:M_comms_SI}

\subsection{$\mbf{\M{glc}}$: aerobic growth under glucose}

This graph has 48 reactions with nonzero flux and 227 edges.  At
Markov time $t=7.66$ (Fig.~\ref{fig:Markov_Stability_fba_networks}A)
this graph has a partition into three communities
(Fig.~\ref{fig:networksM}A):
\begin{itemize}
\item Community \C{1}{\M{glc}} comprises the intake of glucose and
  most of the glycolysis and pentose phosphate pathway. The function
  of the reactions in this community consists of carbon intake and
  processing glucose into phosphoenolpyruvate (PEP). This community
  produces essential biocomponents for the cell such as alpha-D-Ribose
  5-phosphate (rp5), D-Erythrose 4-phosphate (e4p),
  D-fructose-6-phosphate (f6p), glyceraldehyde-3-phosphate (g3p) or
  3-phospho-D-glycerate (3pg). Other reactions produce energy ATP and
  have reductive capabilities for catabolism.

\item Community \C{2}{\M{glc}} contains the electron transport chain
  which produces the majority of the energy of the cell. In the core
  \textit{E coli} metabolic model the chain is represented by the
  reactions NADH dehydrogenase (NADH16), cytochrome oxidase BD (CYTBD)
  and ATP synthase (ATPS4r). This community also contains associated
  reactions to the electron transport such as phosphate intake
  (EXpi(e), PIt2), oxygen intake (EXo2(e), O2t) and proton balance
  (EXh(e)). This community also includes the two reactions that
  represent energy maintenance costs (ATPM), and growth (biomass);
  this is consistent with the biological scenario because ATP is the
  main substrate for both ATPM, and the biomass reaction.
  
\item Community \C{3}{\M{glc}}  contains the TCA cycle at its
  core. The reactions in this community convert PEP into ATP, NADH and
  NADPH. In contrast with \C{1}{\M{glc}}, there is no precursor formation
  here. Beyond the TCA cycle, pyruvate kinase (PYK),
  phosphoenolpyruvate carboxylase (PPC) and pyruvate dehydrogenase
  (PDH) appear in this community. These reactions highlight the two
  main carbon intake routes in the cycle: oxalacetate from PEP through
  phosphoenol pyruvate carboxylase (PPC), and citrate from acetyl
  coenzyme A (acetyl-CoA) via citrate synthase (CS). Furthermore, both
  routes begin with PEP, so it is natural for them to belong to the
  same community along with the rest of the TCA cycle. Likewise, the
  production of L-glutamate from 2-oxoglutarate (AKG) by glutamate
  dehydrogenase (GLUDy) is strongly coupled to the TCA cycle.
\end{itemize}

\subsection{$\mbf{\M{etoh}}$: aerobic growth under ethanol}
This graph contains 49 reactions and 226 edges.  At Markov time
$t=6.28$ (Fig.~\ref{fig:Markov_Stability_fba_networks}B) this graph has a
partition into three communities (Fig.~\ref{fig:networksM}B):
\begin{itemize}
\item Community \C{1}{\M{etoh}} in this graph is similar to its
  counterpart in $\M{glc}$, but with important differences. For
  example, the reactions in charge of the glucose intake (EXglc(e) and
  GLCpts) are no longer part of the network (i.e., they have zero
  flux), and reactions such as malic enzyme NAPD (ME2) and
  phosphoenolpyruvate caboxykinase (PPCK), which now appear in the
  network, belong to this community. This change in the network
  reflect the cell's response to a new biological situation. The
  carbon intake through ethanol has changed the direction of
  glycolysis into gluconeogenesis~\cite{Berg2002} (the reactions in
  \C{1}{\M{glc}} in Fig.~\ref{fig:networksM}A are now operating in the
  reverse direction in Fig.~\ref{fig:networksM}B).  The main role of
  the reactions in this community is the production of bioprecursors
  such as PEP, pyruvate, 3-phospho-D-glycerate (3PG)
  glyceraldehyde-3-phosphate (G3P), D-fructose-6-phosphate (F6P), and
  D-glucose-6-phosphate, all of which are substrates for growth.
  Reactions ME2 and PPCK also belong to this community due to their
  production of PYR and PEP. Reactions that were in a different
  community in $\M{glc}$, such as GLUDy and ICDHyr which produce
  precursors L-glutamate and NADPH respectively, are now part of
  \C{1}{\M{etoh}}. This community also includes the reactions that
  produce inorganic substrates of growth such as NH$_4$, CO$_2$ and
  H$_{2}$O.

\item Community \C{2}{\M{etoh}} contains the electron transport
  chain and the bulk of ATP production, which is similar to
  \C{2}{\M{glc}}. However, there are subtle differences that reflect
  changes in this new scenario. Ethanol intake and transport reactions
  (EXetoh(e) and ETOHt2r) appear in this community due to their
  influence in the proton balance of the cell. In addition, \C{2}{\M{etoh}}
  contains NADP transhydrogenase (THD2) which is in charge of
  NADH/NADPH balance. This reaction is present here due to the NAD
  consumption involved in the reactions ACALD and ethanol
  dehydrogenase (ALCD2x), which belong to this community as well.

\item Community \C{3}{\M{etoh}}  contains most of the TCA cycle. The main
  difference between this community and \C{1}{\M{glc}} is that
  here acetyl-CoA is extracted from acetaldehyde (which comes from
  ethanol) by the reaction acetaldehyde dehydrogenase reaction
  (ACALD), instead of the classical pyruvate from glycolysis. The
  glycoxylate cycle reactions isocitrate lyase (ICL) and malate
  synthase (MALS) which now appear in the network, also belong to this
  community. These reactions are tightly linked to the TCA cycle and
  appear when the carbon intake is acetate or ethanol to prevent the
  loss of carbon as CO$_2$. 
\end{itemize}

\subsection{$\mbf{\M{anaero}}$: anaerobic growth}

\begin{figure}[tp]
 \begin{center}
   \includegraphics[width=\textwidth]{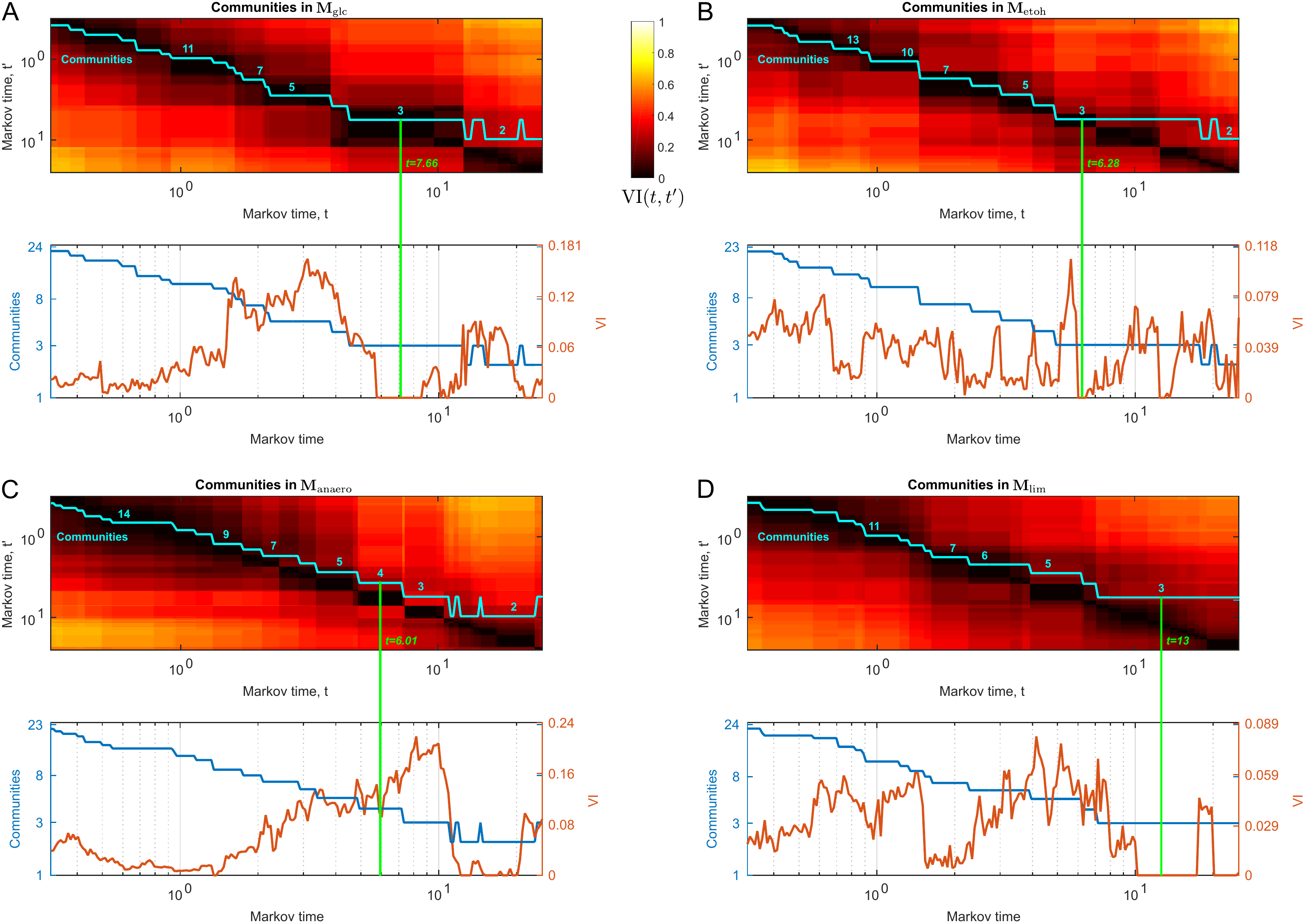}
 \end{center}
 \caption{{\bf Community structure in the MFGs.}  Number of
   communities and VI of the MFGs in four biological scenarios.  (A)
   The graph $\mbf{\M{glc}}$ has a robust partition into three
   communities at $t=7.66$. (B) $\mbf{\M{etoh}}$ has a partition into
   three communities at $t=6.28$. (C) $\mbf{\M{anaero}}$ has four
   communities at $t=6.01$. (D) $\mbf{\M{lim}}$ has three communities
   at $t=13.0$.}
 \label{fig:Markov_Stability_fba_networks}
\end{figure}

This graph contains 47 reactions and 212 edges.  At Markov time
$t=6.01$ (Fig.~\ref{fig:Markov_Stability_fba_networks}C) this graph
has a partition into four communities (Fig.~\ref{fig:networksM}C):
\begin{itemize}
\item Community \C{1}{\M{anaero}} contains the reactions responsible
  D-glucose intake (EXglc) and most of the glycolysis. The reaction
  that represents the cellular maintenance energy cost, ATP
  maintenance requirement (ATPM), is included in this community
  because of the increased strength of its connection to the
  substrate-level phosphorilation reaction phosphoglycerate kinase
  (PGK). Also note that reactions in the pentose phosphate pathway do
  not belong to the same community as the glycolysis reactions (unlike
  in $\M{glc}$ and $\M{etoh}$).

\item Community \C{2}{\M{anaero}} contains the conversion of
  PEP into formate through the sequence of reactions PYK, PFL, FORti
  and EXfor(e). More than half of the carbon secreted by the cell
  becomes formate.
  
\item Community \C{3}{\M{anaero}} includes the biomass reaction and
  the reactions in charge of supplying it with substrates. These
  reactions include the pentose phosphate pathway (now detached from
  \C{1}{\M{glc}}), which produce essential growth precursors such as
  alpha-D-ribose-5-phosphate (r5p) or D-erythrose-4-phosphate
  (e4p). The TCA cycle is present as well because its production of
  two growth precursors: 2-oxalacetate and NADPH. Finally, the
  reactions in charge of acetate production (ACKr, ACt2r and EXac(e))
  are also members of this community through the ability of ACKr to
  produce ATP.  Glutamate metabolism reaction GLUDy is also included
  in this community. It is worth mentioning that the reverse of ATP
  synthase (ATPS4r) is present in this community because here, unlike
  in $\M{glc}$, ATPS4r consumes ATP instead of producing it. When this
  flux is reversed, then ATPS4r is in part responsible for pH
  homeostasis.
  
\item Community \C{4}{\M{anaero}} includes the main reactions involved
  in NADH production and consumption, which occurs via
  glyceraldehyde-3-phosphate dehydrogenase (GAPD). NADH consumption
  occurs in two consecutive steps in ethanol production: in ACALD and
  ALCD2x. The phosphate intake and transport reactions EXpi(e) and
  PIt2r belong to this community because most of the phosphate
  consumption takes place at GAPD. Interestingly, the core reaction
  around which the community forms (GAPD) is not present in the
  community. It is included in earlier Markov times but when
  communities start to get larger the role of GAPD becomes more
  relevant as a part of the glycolysis than its role as a NADH
  hub. This is a good example of how the graph structure and the
  clustering method are able to capture two different roles in the
  same metabolite.
\end{itemize}

\subsection{$\M{lim}$: aerobic growth under limiting conditions}

This graph has 52 nodes and 228 edges. At Markov time $t=13$ this
graph (Fig.~\ref{fig:Markov_Stability_fba_networks}D) has a
partition into three communities (Fig.~\ref{fig:networksM}D):
\begin{itemize}
\item Community \C{1}{\M{lim}} contains the glycolysis pathway
  (detached from the pentose phosphate pathway). This community is
  involved in precursor formation, ATP production, substrate-level
  phosphorylation and processing of D-glucose into PEP.

\item Community \C{2}{\M{lim}} contains the bioenergetic machinery
  of the cell; the main difference to the previous scenarios is that
  the electron transport chain has a smaller role in ATP production
  (ATPS4r), and substrate-level phosphorylation (PGK, PYK, SUCOAS,
  ACKr) becomes more important. In $\M{lim}$ the electron
  transport chain is responsible for the 21.8\% of the total ATP
  produced in the cell while in $\M{glc}$ it produces
  66.5\%. The reactions in charge of intake and transport of inorganic
  ions such as phosphate (EXpi(e) and PIt2r), O$_2$ (EXO$_2$(e) and
  O$_2$t)and H$_{2}$O (EXH$_2$O and H$_2$Ot) belong to this community
  as well. This community includes the reactions in the pentose
  phosphate pathway that produce precursors for growth: transketolase
  (TKT2) produces e4p, and ribose-5-phosphate isomerase (RPI) produces
  r5p.
  
\item Community \C{3}{\M{lim}} is the community that differs the most
  from those in the other aerobic growth networks ($\M{glc}$ and
  $\M{etoh}$). This community gathers reactions that under normal
  circumstances would not be so strongly related but that the limited
  availability of ammonium and phosphate have forced together; its
  members include reactions from the TCA cycle, the pentose phosphate
  pathway, nitrogen metabolism and by-product secretion. The core
  feature of the community is carbon secretion as formate and
  acetate. Reactions PPC, malate dehydrogenase (MDH) reverse and ME2
  channel most of the carbon to the secretion routes in the form of
  formate and acetate. The production of L-glutamine seems to be
  attached to this subsystem through the production of NADPH in ME2
  and its consumption in the glutamate dehydrogenase NAPD (GLUDy).
\end{itemize}

\bibliography{biblio_metabolic}
\bibliographystyle{vancouver}

\end{document}